\documentclass[letterpaper,journal,twoside]{IEEEtran}

\usepackage{blindtext}
\usepackage{graphicx}

\usepackage{graphicx}
\usepackage[table]{xcolor} 
\usepackage{subcaption}
\usepackage{amsmath}
\usepackage{bm} 
\usepackage{amssymb} 
\usepackage{tabularx}
\usepackage{verbatim}
\usepackage[colorinlistoftodos]{todonotes}
\usepackage{balance} 
\usepackage{algorithm} 
\usepackage[noend]{algpseudocode} 

\newcommand{\cablecellcolorbot}{red!15}
\newcommand{\barcellcolorbot}{cyan!15}
\newcommand{\cablecellcolortop}{red!30}
\newcommand{\barcellcolortop}{cyan!40}

\newcommand{\cct}{\cellcolor{\cablecellcolortop}}
\newcommand{\bct}{\cellcolor{\barcellcolortop}}

\DeclareMathOperator*{\argmin}{arg\,min}

\begin{document}


\newcommand{\bq}{\mathbf{q}} 
\newcommand{\dbq}{\dot{\mathbf{q}}} 
\newcommand{\bx}{\mathbf{x}} 
\newcommand{\dbx}{\dot{\mathbf{x}}} 
\newcommand{\bF}{\mathbf{F}} 
\newcommand{\boldf}{\mathbf{f}} 
\newcommand{\bQ}{\mathbf{Q}} 
\newcommand{\bu}{\mathbf{u}} 
\newcommand{\bU}{\mathbf{U}} 
\newcommand{\bolds}{\mathbf{s}} 
\newcommand{\bb}{\mathbf{b}} 
\newcommand{\bB}{\mathbf{B}} 
\newcommand{\br}{\mathbf{r}} 
\newcommand{\bell}{\bm{\ell}} 
\newcommand{\bw}{\mathbf{w}} 
\newcommand{\bv}{\mathbf{v}} 
\newcommand{\bp}{\mathbf{p}} 
\newcommand{\bh}{\mathbf{h}} 
\newcommand{\bzero}{\mathbf{0}} 
\newcommand{\by}{\mathbf{y}} 
\newcommand{\bz}{\mathbf{z}} 
\newcommand{\bE}{\mathbf{E}} 
\newcommand{\be}{\mathbf{e}} 
\newcommand{\ba}{\mathbf{a}} 
\newcommand{\bH}{\mathbf{H}} 
\newcommand{\bJ}{\mathbf{J}} 
\newcommand{\bC}{\mathbf{C}} 
\newcommand{\bOnes}{\mathbf{1}} 
\newcommand{\bI}{\mathbf{I}} 
\newcommand{\bA}{\mathbf{A}} 
\newcommand{\bW}{\mathbf{W}} 
\newcommand{\bL}{\mathbf{L}} 
\newcommand{\bkappa}{\bm{\kappa}} 
\newcommand{\bK}{\mathbf{K}} 
\newcommand{\bg}{\mathbf{g}} 
\newcommand{\bG}{\mathbf{G}} 
\newcommand{\bc}{\mathbf{c}} 
\newcommand{\bX}{\mathbf{X}} 
\newcommand{\bZ}{\mathbf{Z}} 
\newcommand{\dq}{\dot q} 
\newcommand{\brho}{\bm{\rho}} 
\newcommand{\bomega}{\bm{\omega}} 
\newcommand{\balpha}{\bm{\alpha}} 
\newcommand{\bM}{\mathbf{M}} 
\newcommand{\bT}{\mathbf{T}} 
\newcommand{\bR}{\mathbf{R}} 
\newcommand{\bS}{\mathbf{S}} 
\newcommand{\bxi}{\bm{\xi}} 

\bstctlcite{IEEEexample:BSTcontrol}

\title{Model-Predictive Control with Inverse Statics Optimization for Tensegrity Spine Robots}

%



%
        
\author{Andrew P. Sabelhaus,
        Huajing Zhao,
        Edward L. Zhu, \\
        Adrian K. Agogino,
        Alice M. Agogino,
\thanks{This work was supported by NASA Space Technology Research Fellowship no. NNX15AQ55H. ({\it Corresponding author: Andrew P. Sabelhaus.)}}
\thanks{A.P. Sabelhaus and A.M. Agogino are with the Department of Mechanical Engineering, University of California Berkeley, Berkeley CA 94720 USA (email: apsabelhaus@berkeley.edu, agogino@berkeley.edu).}
\thanks{H. Zhao is with the Department of Mechanical Engineering, University of Michigan, Ann Arbor MI 48109 USA (email: hjzhao@umich.edu).}
\thanks{E. Zhu is with the US Army Research Lab, Vehicle Technology Directorate, Aberdeen MD 21005 USA (email: edward.l.zhu.civ@mail.mil).}%
\thanks{A.K. Agogino is with the Intelligent Systems Divison, NASA Ames Research Center, Moffet Field CA 94035 USA (email: adrian.k.agogino@nasa.gov).}}%

\maketitle

\begin{abstract}

Robots with flexible spines based on tensegrity structures have potential advantages over traditional designs with rigid torsos.
However, these robots can be difficult to control due to their high-dimensional nonlinear dynamics and actuator constraints. 
This work presents two controllers for tensegrity spine robots, using model-predictive control (MPC) and inverse statics optimization.
The controllers introduce two different approaches to making the control problem computationally tractable.
The first utilizes smoothing terms in the MPC problem.
The second uses a new inverse statics optimization algorithm, which gives the first feasible solutions to the problem for certain tensegrity robots, to generate reference input trajectories in combination with MPC.
Tracking the inverse statics reference input trajectory significantly reduces the number of tuning parameters.
The controllers are validated against simulations of two-dimensional and three-dimensional tensegrity spines.
Both approaches show noise insensitivity and low tracking error, and can be used for different control goals.
The results here demonstrate the first closed-loop control of such structures.
\end{abstract}



\section{Introduction}


Quadruped (four-legged) robots that are designed with rigid torsos can be limited in the types of terrain they are able to safely traverse \cite{Fukuoka2003,Raibert2008,Buchli2009}.
Alternatively, when quadruped robots are constructed with spine-like flexible bodies that include actuation, significant control challenges are often encountered \cite{Seok2014}.
Few dynamics-based closed-loop control approaches have been developed for robots like these.
Instead, authors commonly use kinematics-only models \cite{Maleki2015,Horvat2015}, model-free control using machine learning \cite{Weinmeister2015,Eckert2015,Hustig-Schultz2016}, decoupled controllers for different parts of the robot \cite{Seok2014}, or the replaying of open-loop inputs \cite{Tsujita2011} among other approaches.


\begin{figure}[htp]
	\centering
	\begin{subfigure}[b]{0.48\columnwidth}
	    \centering
	    \captionsetup{justification=centering}
	    \includegraphics[width=1\columnwidth]{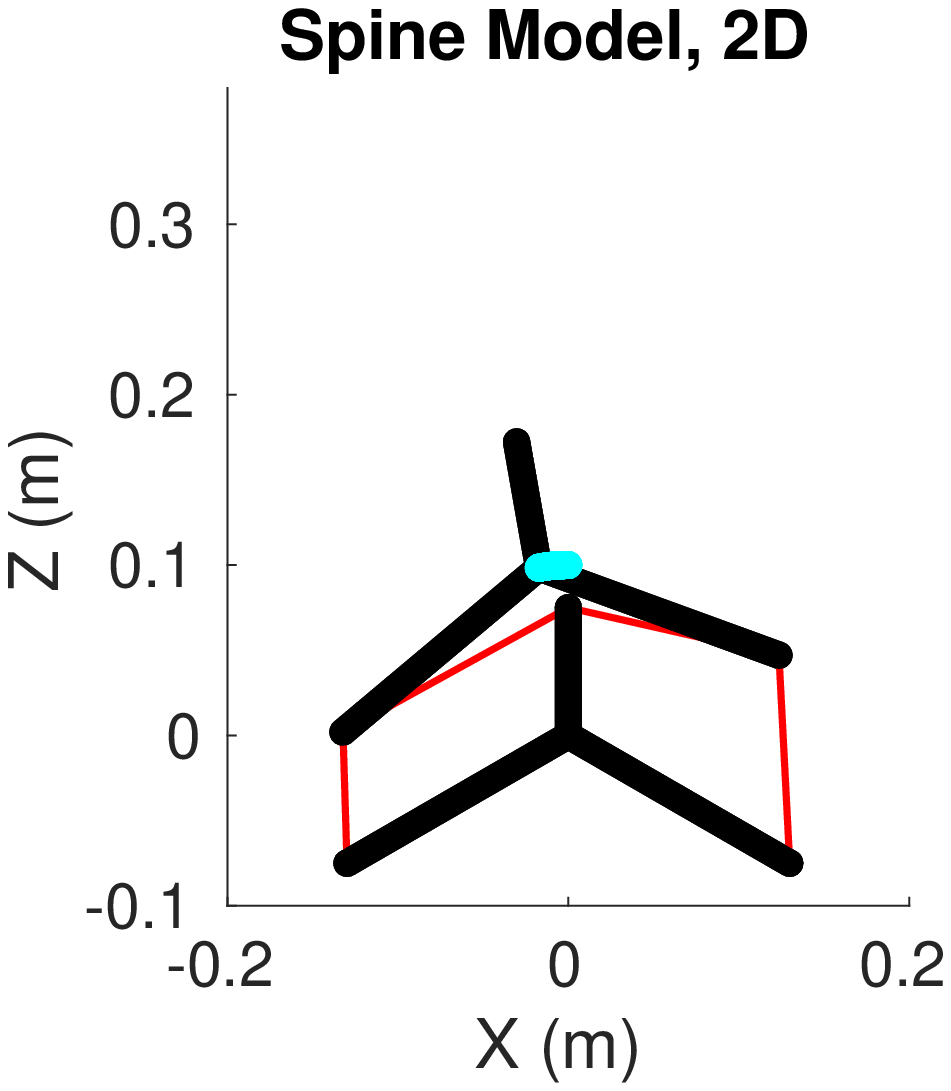}
	    \caption{MPC with inverse statics optimization, 2D model, \\ single moving vertebra.}
	    \vspace{-0.1cm}
	    \label{fig:ultra_spine_mid-bend_mpc_2D}
    \end{subfigure}%
	\begin{subfigure}[b]{0.52\columnwidth}
	    \centering
	    \captionsetup{justification=centering}
	    \includegraphics[width=1\columnwidth]{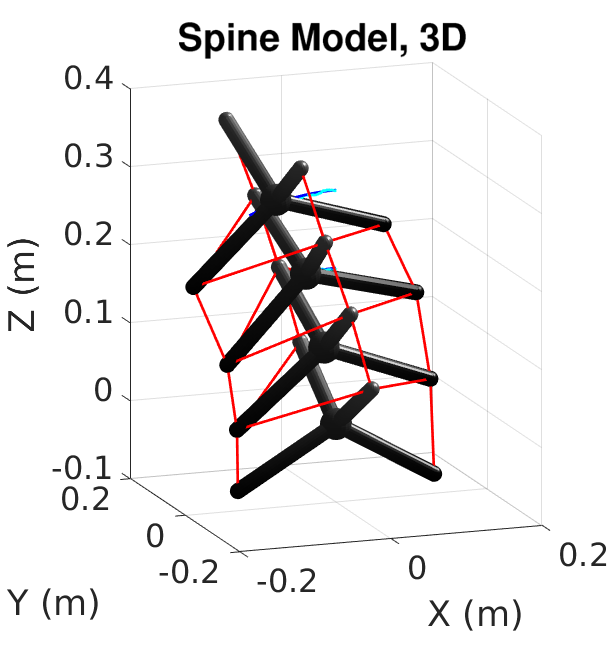}
	    \caption{MPC with smoothing\\ terms, 3D model,\\ three moving vertebrae.}
	    \vspace{-0.1cm}
	    \label{fig:ultra_spine_mid-bend_mpc_3D}
	\end{subfigure}
	\vspace{-0.2cm}
	\caption{Control results for the example tensegrity spine model, for a uniaxial bending trajectory. The rigid bodies (vertebrae) of the spine are in black, cables in red, and the tracked result of the vertebra(e) center-of-mass is in light blue. The bottom-most vertebra is anchored to the ground and does not move, so is not part of the dynamics.}
	\vspace{-0.3cm}
	\label{fig:ultra_spine_mid-bend_both}
\end{figure}


The Laika project, named for the first dog in space, is an ongoing research effort to develop a quadruped robot with a flexible, actuated spine \cite{Sabelhaus2018c}.
Laika's spine is developed as a form of the previously-investigated ULTRA Spine (Underactuated Lightweight Tensegrity Robot Spine) \cite{Sabelhaus2015,Sabelhaus2017}.




\begin{table*}[ht]
\caption{Controller formulations: MPC with smoothing vs. MPC with tracking of inverse statics (IS) input trajectories} 
\label{tab:comparison}
\fontsize{10pt}{12pt}\selectfont
\vspace{-0.2cm}
\begin{center}%
\begin{tabular}{|l|lllll|}
    \hline
    Controller formulation & \# Tuning constants & Time discr. & Simulation setup & Max. Error & Refs. \\ \hline \hline
    Smoothing terms & 14 & $1e^{-3}$ sec. & 3 vertebra, 3D & $<0.5$ cm & \cite{Sabelhaus2017} \\
    IS input traj. tracking & 5 & $1e^{-5}$ sec. & 1 vertebra, 2D & See Sec. \ref{sec:results_invstat} & - \\
    \hline
\end{tabular}
\end{center}
\smallskip
The two controller formulations presented in this paper have different benefits with respect to tuning and performance. The five tuning constants (column two) of the more general controller with the inverse statics optimization algorithm are straightforward to chose. 
All have physical interpretations (e.g., minimum cable tension, vertebra anti-collision distance) or are common to many optimal control problems (e.g., the $\bQ$ and $\bR$ weighting matrices in eqns. (\ref{eq:Q_invstat})-(\ref{eq:R_invstat}), and MPC horizon length.)
\vspace{-0.1cm}
\end{table*}

This work presents two controllers for Laika's spine, both of which track state-space trajectories using the spine's dynamics model.
These controllers use combinations of model-predictive control (MPC) with a new inverse statics (IS) algorithm.
Both frameworks are motivated by the practical challenges with computational complexity of nonlinear, optimization-based control.
The first controller, presented in the conference version of this work \cite{Sabelhaus2017}, demonstrated proof-of-concept by employing a variety of smoothing and tuning terms in the MPC optimization problem.

The second controller incorporates an inverse statics optimization problem to generate reference input trajectories that are then used with MPC.
The new approach is significantly more general, with less tuning than the smoothing approach (Table \ref{tab:comparison}), and with more favorable computational characteristics since the inverse statics are solved offline.
Along the way, a new algorithm is derived for the inverse statics of tensegrity structures with internal bending moments.
This approach contributes a new architecture for addressing computationally-complex state tracking problems in robots such as these spines.
The controller's novelty arises from both this new solution to the inverse statics problem as well as its interconnection with MPC, forming a new approach to control of such systems (Sec. \ref{sec:ctlr_invstat} and Fig. \ref{fig:MPC_IS_block_diagram}).

This work uses a three-dimensional model of the spine, with multiple vertebrae, for evaluating the smoothing controller (from \cite{Sabelhaus2017}.)
Meanwhile, a reduced-order two-dimensional model, with only a single moving vertebra, is used for the controller with the IS optimization for input trajectory generation.
Although the controller with the IS optimization is tested on a lower-dimensional system than the smoothing controller, the tracked states of the vertebrae are the same.
Therefore, the results are compared quantitatively in Sec. \ref{sec:results}, and the limitations of this comparison are discussed in Sec. \ref{sec:discussion}.

\section{Background}



The spine robot studied in this work (Fig. \ref{fig:ultra_spine_mid-bend_both}) is a tensegrity, or ``tension-integrity'', structure.
Tensegrity structures consist of rigid bodies suspended in a network of cables in tension such the bodies do not contact each other \cite{skelton2009tensegrity}. 
This definition is sometimes restricted further to structures without internal bending moments, i.e., where the bodies are single bars and cables only connect at bar ends \cite{Moored2006,Tur2009,Bliss2012a}.
However, both historical examples (Snelson's `X-Cross', \cite{skelton2009tensegrity}) and many modern robots \cite{Mirletz2015a,Friesen2016,Bohm2016a,Friesen2018,Chen2019} use the broader definition, having more complicated bodies suspended in the network, as with the vertebrae of these spines.

Tensegrity structures are inherently flexible, and many types of tensegrity robots have been designed that leverage this flexibility. These robots are able to adjust the lengths of their cables to roll \cite{Koizumi2012b,Sabelhaus2015a,Kim2015,Chen2016,Rieffel2018,Vespignani2018a}, crawl \cite{Paul2006a,Shibata2009b,Tietz2013,Mirletz2015a,Zappetti2017}, swim \cite{Bliss2013a,Chen2019}, hop and jump \cite{Kim2016,Mintchev2018}, and climb \cite{friesen2014,Friesen2016}.
Tensegrity spine robots have been previously investigated \cite{mirletz2014design,Mirletz2015,Mirletz2015a}, but the ULTRA Spine and its recent adaptation for Laika are one of the first uses of a tensegrity spine on a quadruped robot \cite{Sabelhaus2015,Hustig-Schultz2016}.

\vspace{-0.1cm}
\subsection{Control of Tensegrity Structures and Robots}\label{sec:background_tensegrity_control}


Although there are a variety of benefits to using tensegrity structures as robots, control of such structures has proven challenging.
This is commonly due to dynamics which are inherently nonlinear and often high-dimensional.
Various saturation issues also exist, as cables within the structure exert no force in compression, and the controller cannot retract the spine's flexible cables to a negative `rest length' (defined in Sec. \ref{sec:cable_model}.)
Consequently, state-space control for tracking or regulation has been mostly limited to low-dimensional structures, particularly those with only bars \cite{Aldrich2003,Wroldsen2009,skelton2009nonlinear,skelton2009tensegrity,Tur2009}, which assume, a-priori,  that all cables are initially tensioned.
Open-loop methods have also been used for this purpose \cite{Sultan2002,Sultan2003,Caluwaerts2015a,friesen2014}, but cannot reject disturbances.

However, successful control strategies have been developed for tensegrity robots for other control goals.
In particular, when the robots are intended to roll or crawl, model-free controllers have used evolutionary algorithms \cite{Paul2006a,Iscen2014a,Iscen2015,Khazanov2014}, central pattern generators \cite{Mirletz2015,Mirletz2015a}, Bayesian optimization \cite{Rieffel2018,Kimber2019}, deep reinforcement learning \cite{Zhang2017}, kinodynamic motion planning \cite{Littlefield2019}, or hand-tuned algorithms \cite{Vespignani2018}.
Model-predictive control has been used for generating locomotion primitives for imitation learning \cite{Cera2018}.
Tensegrity structures which oscillate, such as fish tails, have used resonance entrainment \cite{Bliss2012a,Bliss2013a}.
Though these approaches are promising in their domains, they do not necessarily apply to state tracking, as is needed here.


\subsection{Inverse Statics and Form-Finding for Robotics Control}\label{sec:background_invstat}



Though control of tensegrity robots is challenging, the related problem of \textit{form-finding} has a variety of well-known solutions \cite{Tibert2003}.
Form-finding simultaneously solves for a pose of the tensegrity structure's bodies alongside the cable forces that keep it in equilibrium.
In the context of control, these solutions correspond to equilibrium setpoints for both states and inputs.
A subset of this problem is solving for the cable forces in static equilibrium for a given pose, i.e., solving for inputs given a desired state.
For related parallel robots, the former problem is termed \textit{inverse kinetostatics} analysis \cite{Birglen2004,Belfiore2013}, and the latter subset is \textit{inverse statics} analysis \cite{Arsenault2006,Giorelli2015}.

Solutions to the inverse statics problem have been used as part of open-loop controllers \cite{friesen2014,Camarillo2009,Giorelli2015}, or more rarely closed-loop controllers \cite{BarretoS.2014}, of a variety of robots.
Inverse statics is simpler to formulate than inverse dynamics, has computational benefits since optimization problems can be solved offline \cite{friesen2014,Camarillo2009}, and provides a reasonable approximation to dynamic trajectories under pseudo-static movements.
For tensegrity robots like these spines, no kinematic constraints exist between the bodies.
Since the robot's statics and kinematics are therefore decoupled, kinetostatic approaches such as \cite{Gao2014a,Arsenault2006,Belfiore2013} are not required.
This allows for the easier-to-solve \textit{force-density} method \cite{Schek1974,Tibert2003,Tran2010} to be applied.
However, solutions have yet to exist in the literature for tensegrity robots with internal bending moments, of the type considered here.




\subsection{Model-Predictive Control for High-Dimensional Nonlinear Robots and Systems}\label{sec:background_mpc}


This work proposes controllers based on model-predictive control (MPC) for three primary reasons.
First, using an optimization program for control can address constraints on the system (actuator saturation and tensioned cables). 
Second, computational tractability can be addressed by using a receding horizon.
These two features define an MPC problem.
Finally, an MPC formulation allows straightforward introduction of smoothing weights and constraints for hard-to-control systems \cite{Worthmann2016,Falcone2007}.
These motivated the original MPC formulation in the conference version of this work \cite{Sabelhaus2017}.

Model-predictive control for nonlinear systems (NMPC) is a well-studied topic with many implementations \cite{allgower2000nonlinear}, particularly in low dimensions where nonconvex optimization is computationally feasible \cite{Worthmann2016}.
For high-dimensional nonlinear systems, practical options include modifying the NMPC problem or using more efficient solvers \cite{Cannon2004}.
Alternatively, linearized dynamics can be used at later points in the horizon \cite{Zheng1997}, or linearizations can be performed at each timestep in the problem to create a linear time-varying MPC \cite{Falcone2007}.
A time-varying linearization is used in this work for computational tractability purposes.
As opposed to attempting to solve an NMPC problem in real time for this high-dimensional system, the proposed control architecture addresses linearization error via two other methods. 
In particular, one approach here includes smoothing terms, while the other tracks an approximated input reference trajectory generated by inverse statics.

\vspace{-0.1cm}
\subsection{Simulation-Based Controller Validation}\label{sec:background_simulation}



This work employs a set of simulations that demonstrate the performance of the proposed controllers and show proof-of-concept.
There are fewer traditional sources of modeling error in this problem than in other tensegrity robotics control problems, which commonly involve locomotion on the ground \cite{Paul2006a,Zhang2017}. 
Locomotion requires modeling complex interactions with the ambient environment \cite{Bliss2013a}. 
This spine instead moves freely in space without surface contact, therefore simulation inaccuracies due to contact friction modeling are not present. 
Here, the most significant sources of error are anticipated to arise from unmodeled actuator dynamics and manufacturing differences of hardware prototypes versus the nonlinear dynamics model.
Prior work has confirmed that rigid body models of free-standing tensegrity robots match hardware results reasonably well under similar conditions \cite{Bliss2012,Caluwaerts2014}.

Though the simulation setup captures the core dynamics phenomena of the system, a future physical prototype would confirm the magnitude of the `reality gap' between simulation and hardware.
There remain numerous technical challenges to doing so.
In particular, there are significant actuation and sensing challenges for a prototype of this complexity, and in addition, the controllers presented here are not yet real-time. 
Future work seeks to contribute new hardware designs as well as even more computationally-efficient controllers so that validation and verification experiments can be performed.


\section{Spine Model and Movement Goals}\label{sec:dynamics_section}


The geometry of the spine and its equations of motion are adapted from earlier work \cite{Sabelhaus2017}. 
The following section briefly describes the state-space model used for both the two-dimensional and three-dimensional spine, as well as the desired state trajectory to be tracked.
Full knowledge of the system states at each timestep is assumed; the controllers in this work are state-feedback.

\vspace{-0.1cm}
\subsection{Vertebra Geometry and State Space}\label{sec:topology}

\begin{figure}[thpb]
    \centering
    \begin{subfigure}{1.0\columnwidth}
        \centering
        \includegraphics[width=0.7\columnwidth]{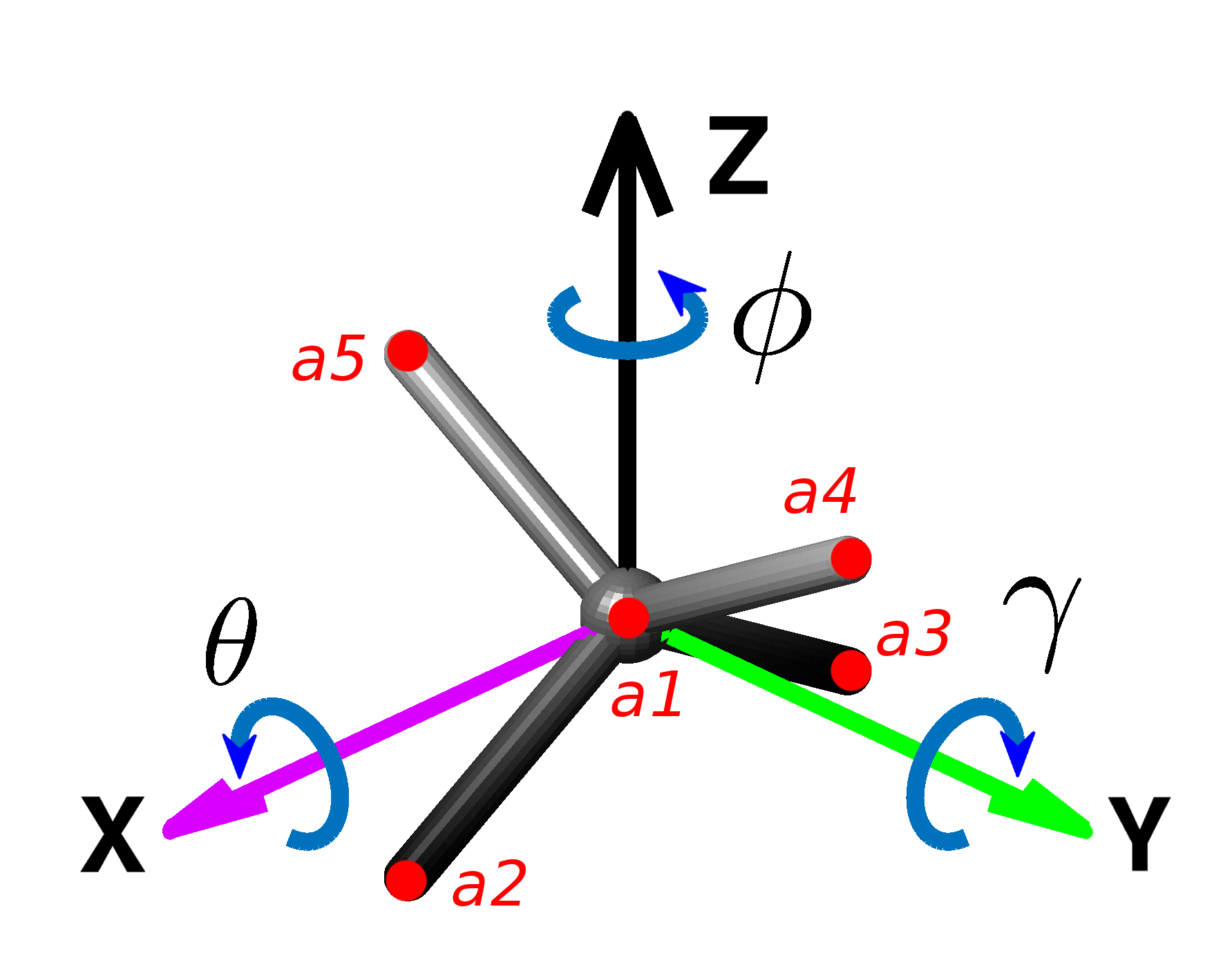}
        \caption{A single 3D spine vertebra in its local coordinate system.}
        \label{fig:topology_3D}
    \end{subfigure}
    \begin{subfigure}{1.0\columnwidth}
        \centering
        \includegraphics[width=0.7\columnwidth]{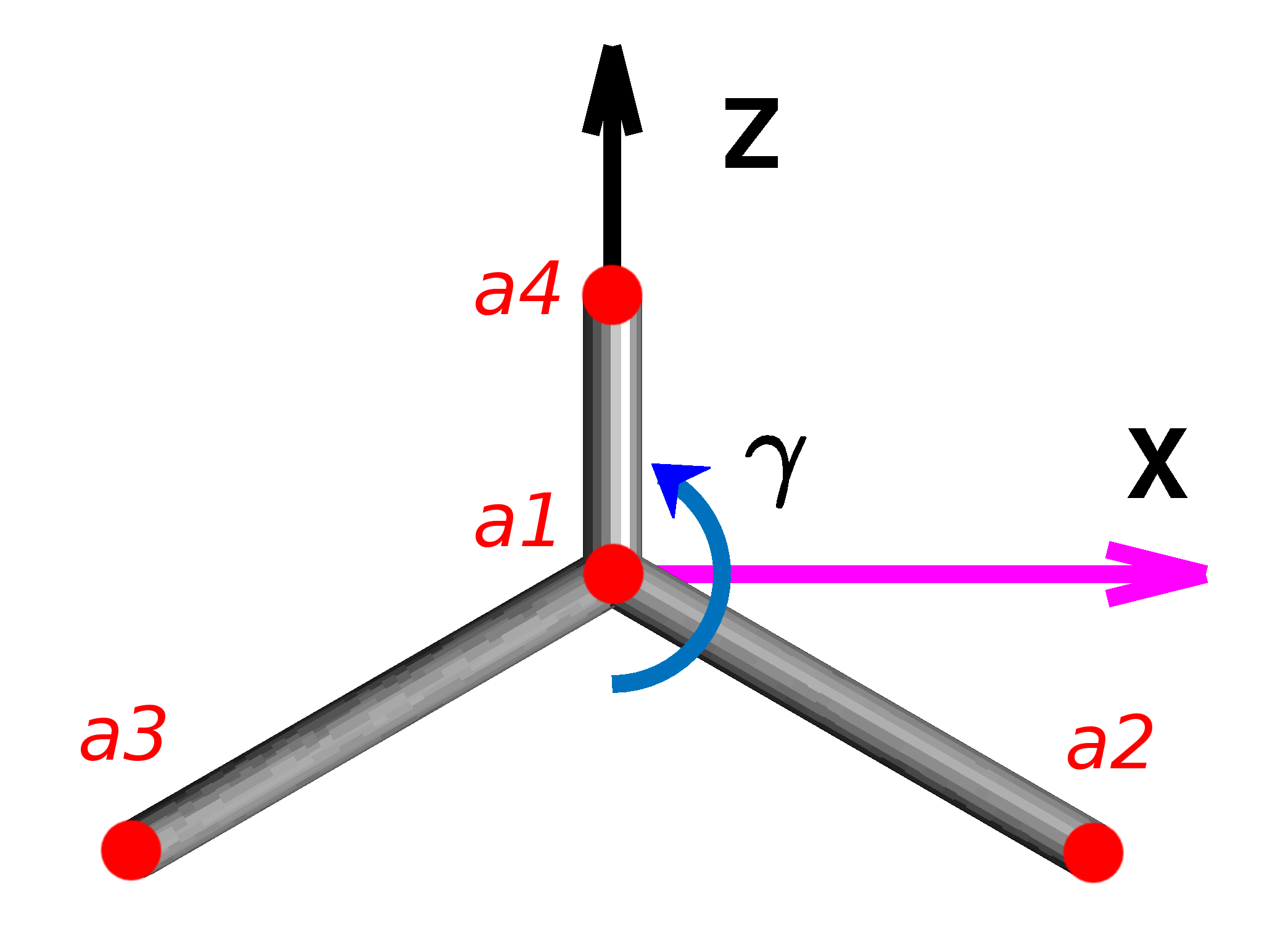}
        \caption{A single 2D spine vertebra in its local coordinate system.}
        \label{fig:topology_2D}
    \end{subfigure}
    \caption{Geometry of spine vertebrae in both 3D and 2D. Point mass locations $\{ \ba_1...\ba_4, \ba_5 \}$ shown in red. Certain coordinates ($X$-axis and $\theta,\gamma$ rotations) are flipped from the right-hand convention in order to match a simulation environment used in prior research \cite{friesen2014,Sabelhaus2015,Sabelhaus2018c}.}
    \label{fig:topology}
    \vspace{-0.4cm}
\end{figure}

Each spine vertebra is a rigid body, approximated by a system of point masses (Fig. \ref{fig:topology}), as has been justified in past literature \cite{Caluwaerts2014,friesen2014}.
The local frame of one vertebra contains point mass $k$ at position $\ba_k$, $k=1\hdots K$ (Fig. \ref{fig:topology}.)

The robot's state space is parameterized by the coordinates of the center of mass and a set of Euler angles (3-2-1) for each vertebra, in addition to their respective time-derivatives.
The continuous-time equations of motion have the form

\vspace{-0.2cm}
\begin{equation}\label{eq:ss_definition}
    \dot \bxi = \bg(\bxi, \bu),
\end{equation}

\noindent
where $\bxi \in \mathbb{R}^{36}$ in 3D (for three moving vertebrae) or $\mathbb{R}^{6}$ in 2D (for one moving vertebra) is the state vector, and $\bu \in \mathbb{R}^{24} \text{ in 3D or } \mathbb{R}^{4} \text{ in 2D}$ is the input vector, which has the same dimension as number of cables. 
Eqn. (\ref{eq:ss_definition}) is expressed using Lagrange's equations, and a symbolically-solved solution is discussed in Appendix Sec. \ref{appendix:eqns_of_motion}.

\vspace{-0.5cm}


\subsection{Cable Model as System Inputs}\label{sec:cable_model}

The cables suspending the vertebrae provide the control input to the system.
Unlike work such as \cite{Wroldsen2009,skelton2009nonlinear,Aldrich2003}, it is not assumed that the controller specifies forces in the robot's cables, since this becomes challenging to implement on physical hardware.
Instead, the control inputs are the rest lengths of a virtual spring-damper. 
Specifically, let the vector between the two connection points of cable $i$ be $\bell_i$, with scalar length $\ell_i = ||\bell_i||$.
Then the applied force due to a cable, directed away from its attachment point, is



\vspace{-0.1cm}
\[
\bF_{i} = F_i(\ell_i, \dot \ell_i) \hat \bell_i,
\]

\noindent so that tension forces are positive.
Here $\hat \bell_i$ is a normalized direction vector.
The scalar tension force on cable $i$ is a rectified spring-damper, so the cables apply no compression forces:

\vspace{-0.2cm}
\begin{equation}\label{eq:springcable}
F_i = 
\begin{cases}
k( \ell_i - u_{i}) - c \dot{\ell_i }, & \text{if} \quad k( \ell_i - u_{i}) - c \dot{ \ell_i } \geq 0 \\
0, & \text{if} \quad k( \ell_i - u_{i}) - c \dot{ \ell_i } < 0, \\
\end{cases}
\end{equation}


\noindent where the input $u_i$ is the rest length of cable $i$.
In addition, the controller cannot command a negative $u_i$, since retracting a cable to a negative length is not physically possible.


\vspace{-0.1cm}
\subsection{Reference State Trajectory}\label{sec:state_trajectories}

\begin{figure}[bth]
    \centering
    \vspace{-0.1cm}
    \includegraphics[width=0.75\columnwidth]{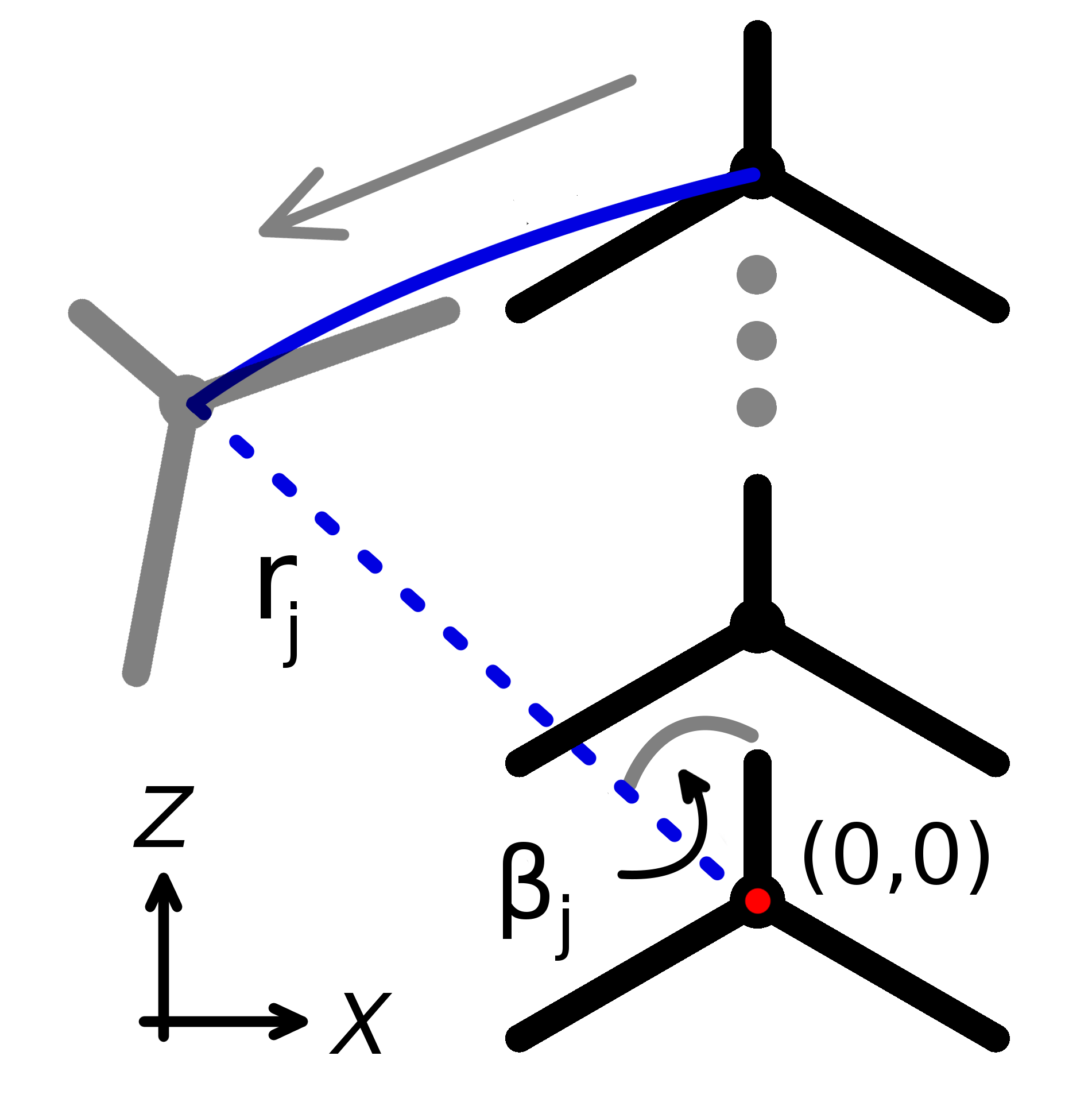}
    \caption{Bending trajectory for the $j$-th vertebra of the spine in the $X$-$Z$ plane. The vertebra rotates counterclockwise around the origin at a constant radius $r_j$ (dashed blue line), swept out by angle $\beta_j$ (solid gray line). Solid blue line shows the center of mass of the vertebra. Details given in Appendix Sec. \ref{appendix:state_traj}.}
    \label{fig:spine_traj}
    \vspace{-0.2cm}
\end{figure}

In this work, the desired trajectory $\bar \bxi$ for the spine robot is a bending motion in the $X$-$Z$ plane, consisting of translations and rotations for each moving vertebra (Fig. \ref{fig:spine_traj}.)
As no a-priori dynamic trajectories were available for this model, the controllers in Sec. \ref{sec:controller} do not include the tracking of vertebrae velocities.
Consequently, this trajectory is not guaranteed to be dynamically feasible. 
However, this sequence of states has been observed as the output of prior qualitative simulation studies in \cite{Sabelhaus2015}, and is therefore judged as a reasonable control goal.





\section{Inverse Statics Optimization for Input Reference Trajectory Generation}\label{sec:invstat}

The second of the two controllers presented in this work includes an algorithm to calculate a reference input trajectory for the spine.
In general, model-based controller formulations for trajectory tracking require both state and input trajectories, i.e. $\bar \bxi(t)$ and $\bar \bu(t)$, at each timestep in discrete time (\cite{astrom2008feedback}, ch. 7.5).
Generation of such a $\bar \bu$ is challenging for nonlinear systems.
The proposed controller therefore uses an approximation: an inverse statics solver is used to find a $\bar \bu$ from $\bar \bxi$, as opposed to a more computationally and practically challenging-to-implement inverse dynamics solver.
Though the resulting combination of these $\bar \bxi$ and $\bar \bu$ is not dynamically feasible, it is close to the feasible solution for slow or pseudo-static movements of the spine. 

The following section presents the inverse statics optimization algorithm, and shows that the trajectory generation solution can be obtained offline via a quadratic program.
The algorithm is adapted from the well known \textit{force-density method} \cite{Schek1974} to allow its application to this spine model.
This reformulation introduces the first feasible inverse statics solutions for tensegrity structures with internal bending moments, such as this spine.






\vspace{-0.1cm}
\subsection{Force Density Method for Tensegrity Networks}\label{sec:forcedensity_graph}

The force density method calculates the static equilibrium condition for cable networks \cite{Schek1974}.
It has, by extension, been used for tensegrity systems as networks of force-carrying structural members in tension or compression \cite{Tran2010}.
This section briefly derives the static equilibrium condition for a structure using force density, as it is required for the proposed algorithm in later subsections.

The equilibrium condition is constructed using the force density method via a force balance at each node in the network.
It therefore assumes that the structure can be represented by a graph of nodes and connections between those nodes, that forces are only present at nodes, and that members do not have mass nor deform.
This implicitly assumes that no moments are present at nodes, which is not appropriate for this spine, and which motivates the reformulation in Sec. \ref{sec:reformulation}.




To calculate the equilibrium condition, assume the tensegrity structure has $n$ nodes.
Let the tensegrity exist in either a $d=2$ or $d=3$ -dimensional physical space.
In two dimensions, denote the coordinates of the nodes, and the external applied forces (not due to cables) at the nodes, as

\vspace{-0.2cm}
\begin{equation}
    \bx, \; \bz, \; \bp_x, \; \bp_z \quad \in \mathbb{R}^n.
\end{equation}

\noindent For the spine, $\bx$ and $\bz$ are the coordinates of the point masses in the global frame, obtained by transforming each vertebra's node $\ba_k$ from Fig. \ref{fig:topology_2D} according to $\bar \bxi(t)$.
External forces $\bp_z$ are due to gravity, $-mg$, at each node.
The later reformulation removes the nodal force balance at the fixed vertebra, eliminating the need to calculate its reaction forces nor add those forces to $\bp_x$ and $\bp_z$.

Next, let the structure have $s$ members in tension (cables) and $r$ members in compression (bars, or parts of a rigid body).
A connectivity matrix $\bC \in \mathbb{R}^{(s+r) \times n}$ can be written that describes how nodes are connected by cable and bar members, where the first $s$ rows of $\bC$ are assumed to correspond to cable members and the last $r$ rows correspond to bar members.
This matrix $\bC$ is defined using a graph structure, where if member $i \in \{1,...,(s+r)\}$ connects nodes $k$ and $j$, then the $k$-th and $j$-th columns in $\bC$ are set to 1 and -1 respectively for row $i$, as in



\vspace{-0.1cm}
\begin{equation}
    \bC^{(i,k)} = 1, \quad \quad \bC^{(i,j)} = -1.
\end{equation}
\vspace{-0.2cm}

\noindent All other entries in $\bC$ are 0.


One of the proposed controllers considers the two-dimensional ($d=2$), single-moving-vertebra version of the spine (Fig. \ref{fig:ultra_spine_mid-bend_mpc_2D}).
Its connectivity matrix is the following, highlighted according to its block structure,





\begin{equation}\label{eq:connectivity}
\bC
=
\left[ \begin{array}{cccccccc}
  \rowcolor{\cablecellcolorbot} 0 &  1 &  0 &  0 & \cct 0 & \cct -1 & \cct 0 & \cct 0 \\
  \rowcolor{\cablecellcolorbot} 0 &  0 &  1 &  0 & \cct 0 & \cct 0 & \cct -1 & \cct 0 \\
  \rowcolor{\cablecellcolorbot} 0 &  0 &  0 &  1 & \cct 0 & \cct -1 & \cct 0 & \cct 0 \\
  \rowcolor{\cablecellcolorbot} 0 &  0 &  0 &  1 & \cct 0 & \cct 0 & \cct -1 & \cct 0 \\
  \rowcolor{\barcellcolorbot} 1 & -1 &  0 &  0 &  \bct 0 & \bct 0 & \bct 0 & \bct 0 \\
  \rowcolor{\barcellcolorbot} 1 &  0 & -1 &  0 &  \bct 0 & \bct 0 & \bct 0 & \bct 0 \\
  \rowcolor{\barcellcolorbot} 1 &  0 &  0 & -1 &  \bct 0 & \bct 0 & \bct 0 & \bct 0 \\
  \rowcolor{\barcellcolorbot} 0 &  0 &  0 &  0 &  \bct 1 & \bct -1 & \bct 0 & \bct 0 \\
  \rowcolor{\barcellcolorbot} 0 &  0 &  0 &  0 &  \bct 1 & \bct 0 & \bct -1 & \bct 0 \\
  \rowcolor{\barcellcolorbot} 0 &  0 &  0 &  0 &  \bct 1 & \bct 0 & \bct 0 & \bct -1
\end{array} \right],
\end{equation}

\vspace{0.3cm}

\noindent where the red rows represent the $s=4$ cables and blue rows are the $r=6$ bars within the vertebrae.
Columns 1-4 (lighter) correspond to the nodes for the bottom vertebra which is fixed to the ground, and 5-8 (darker) correspond to the moving vertebra.



Finally, define the \textit{force density} vector $\bq$ as

\begin{equation} 
\bq = [ \; q_1, \; q_2, \; q_3, \; ... \; q_{s+r} \; ]^\top \in \mathbb{R}^{(s+r)},
\end{equation}


\noindent such that if member $i$ experiences a force $F_i$ along its length of $\ell_i$, 

\begin{equation}\label{eq:forcedensity_defn}
    q_i = F_i/\ell_i.
\end{equation}

\noindent As seen in \cite{friesen2014,Schek1974,Tibert2003,Tran2010} the force balance condition for static equilibrium of the structure can then be stated as

\begin{equation}\label{eq:statics_eq}
\begin{aligned}
    \bC^\top \text{diag}(\bq) \bC \bx = \bp_x, \\
    \bC^\top \text{diag}(\bq) \bC \bz = \bp_z.
\end{aligned}
\end{equation}

\noindent As also discussed in \cite{friesen2014,Schek1974,Tran2010}, eqn. (\ref{eq:statics_eq}) can be reorganized as 

\vspace{-0.2cm}
\begin{equation}\label{eq:Aqp_forcedensity}
    \bA \bq = \bp,
\end{equation}
\vspace{-0.65cm}


\vspace{-0.2cm}
\begin{gather}
\bA = 
\begin{bmatrix}
    \bC^\top \text{diag}(\bC \bx) \\
    \bC^\top \text{diag}(\bC \bz)
\end{bmatrix} \in \mathbb{R}^{(nd) \times (s+r) }, \label{eq:A_forcedensity}
\\
\bp = 
\begin{bmatrix}
    \bp_x \\
    \bp_z
\end{bmatrix} \in \mathbb{R}^{(nd)}. \label{eq:p_forcedensity}
\end{gather}

Here, $\bA$ and $\bp$ are constants at timestep $t$, since $\bx$ and $\bz$ are only a function of $\bar\bxi(t)$ at that timestep.
Therefore, eqn. (\ref{eq:Aqp_forcedensity}) is a set of linear equations in $\bq$.
A value for $\bq$ that satisfies (\ref{eq:Aqp_forcedensity}) can then be obtained in a variety of ways, e.g. by a quadratic program \cite{friesen2014}, which produces a set of equilibrium cable forces for a desired pose.
If solutions exist, then $\bar \bu$ can be calculated from eqns. (\ref{eq:springcable}) and (\ref{eq:forcedensity_defn}) as

\begin{equation}\label{eqn:backcalculate_u}
    \bar u_i = \ell_i - \frac{\ell_i q_i^*}{k_i}.
\end{equation}

\subsection{Existence of Solutions to the Inverse Statics Problem}

For the tensegrity spine robots in this research, no solutions exist to eqn. (\ref{eq:Aqp_forcedensity}).
For the 2D spine with $\bC$ from eqn. (\ref{eq:connectivity}), the $\bA$ matrix is taller than it is wide, with $(s+r) = 10$ and $(nd) = 16$, so has an empty null space in almost all poses.
In works which have previously used the force density method, such as \cite{friesen2014,Sabelhaus2015,Friesen2016}, the tensegrity structure has many more cables and bars than nodes, such that $(s+r) > (nd)$.
In addition to this rank issue, the spine in this work intuitively requires internal bending moments to be present in static equilibrium.
Consider, for example, the center node of the moving vertebra.
Therefore, even if $\bA$ had a nonzero null space, eqn. (\ref{eq:Aqp_forcedensity}) would likely be inconsistent.




This rank deficiency issue for static equilibrium is discussed in the literature on tensegrity structures in the context of geometry \cite{Calladine1978} and energy methods \cite{Connelly1992}.
Algorithms exist for determining if a structure would have static equilibrium solutions \cite{Pellegrino1986,Tur2009}.
However, addressing this issue usually consists of adding cables or changing the geometry of the tensegrity structure itself (via form-finding, e.g. \cite{Tran2010}), which is not possible given the problem statement in this work.

No force-density formulations of static equilibrium for tensegrity structures with internal bending moments are present in the literature.
The following section contributes the first formulation for this case, which produces the first inverse statics solutions for tensegrity robots such as these.


\vspace{-0.1cm}
\subsection{Rigid Body Reformulation of the Force Density Method}\label{sec:reformulation}



This work adapts the node-graph formulation of the force density method into a force balance per rigid body, adds a moment balance for the system, and produces a new set of linear equations for static equilibrium.
Doing so neglects the internal stresses within the vertebrae, consistent with the assumption of rigid-body equations of motion.
The process below is therefore described as a ``rigid body reformulation'', although prior statics work uses the term `rigid' in different contexts \cite{Calladine1978}.





The following derivation, specific to the two-body, two-dimensional tensegrity robot in this work, requires the following two assumptions:

\vspace{0.1cm}
\begin{enumerate}
    \item The tensegrity robot consists of $b$ rigid bodies each with the same number of nodes, $\eta = n/b$.
    \item The columns of $\bC$ are block-ordered according to rigid body: nodes are assigned an ordering in blocks of $\eta$.
\end{enumerate}

\vspace{0.1cm}

\noindent These assumptions are demonstrated in the highlighted $\bC$ in eqn. (\ref{eq:connectivity}).
The robot has $b=2$ bodies, with $\eta = 4$ nodes each, so that columns 1-4 and 5-8 correspond to each body.
This is similar to the repeated `cells' of a larger tensegrity, as the term is used in \cite{Motro2003}. \\


\subsubsection{Force balance per rigid body}

The nodal force balance, eqn. (\ref{eq:Aqp_forcedensity}), can be converted into a force balance per-body via the following.
First, consider eliminating the constraints associated with the stationary body: this allows for arbitrary reaction forces at the body's nodes, adding degrees of freedom to the problem.
Such an assumption is consistent with the body being rigidly fixed to the ground, and takes advantage of the static indeterminacy of the structure.
Eliminating those nodes can be done by left-multiplying a matrix $\bW_f$ to eqn. (\ref{eq:Aqp_forcedensity}), which removes the corresponding rows from $\bA$: 

\vspace{-0.1cm}
\begin{equation}\label{eqn:W_defn}
    \bW = [\bzero_{\eta \times \eta} \; \; \bI_{\eta}], \quad \quad \bW_f = \begin{bmatrix} 
        \bW & \bzero \\ 
        \bzero & \bW
        \end{bmatrix}.
\end{equation}

\noindent The shape of $\bW_f$ arises from the force balance in both the $\bx$ and $\bz$ directions: compare to the block structure in eqn. (\ref{eq:A_forcedensity}).

Next, the rows corresponding to the remaining body can be collapsed with another left-multiplication by 

\vspace{-0.1cm}
\begin{equation}
    \bK = \bI_d \otimes \bOnes_\eta^\top,
\end{equation}

\noindent which combines the per-node balance (each row) for the moving vertebra in both the $\bx$ and $\bz$ directions.
Finally, a right-multiplication of $\bA$ by the following matrix $\bH$ eliminates the constraints associated with the bars, which no longer have physical meaning.
Specifically,

\vspace{-0.1cm}
\begin{equation}\label{eqn:H_defn} 
\bH = \begin{bmatrix}
    \bI_{s} \\
    \bzero_{r \times s}
\end{bmatrix} \in \mathbb{R}^{(s+r) \times s},
\quad \quad
\bH \bq_s = \begin{bmatrix} \bq_s \\ \bzero_r \end{bmatrix}.
\end{equation}



\noindent The identity block in $\bH$ is sized according to the number of cables in the structure, $s$, and corresponds to the rows of eqn. (\ref{eq:connectivity}) that are highlighted in red.
Combining these operations produces

\vspace{-0.3cm}
\begin{equation}
    \bA_f = \bK \bW_f \bA \bH.
\end{equation}

Similarly, the external forces at each node are combined per-body, removing the stationary vertebra and collapsing the constraints in both directions:

\vspace{-0.1cm}
\begin{equation}
    \bp_f = \bK \bW_f \bp.
\end{equation}

\noindent The final result is a per-body force balance for the moving vertebra alone.
Since the bar members have been removed, the free variable only consists of the cable force densities, which are the first $s$ elements of $\bq$, i.e. $\bq_s$.
The constraint is then

\vspace{-0.3cm}
\begin{equation}\label{eq:Aqp_rigidbody}
    \bA_f \bq_s = \bp_f.
\end{equation}




\subsubsection{Moment balance per rigid body}

A moment balance for the moving vertebra is now required.
Moments due to all forces can be summed around any point in the structure (not necessarily the centers of mass) in static equilibrium.
A convenient point is therefore the origin, so that the moment arms are simply the nodal coordinates.
Moments can then be expressed using matrix multiplication as the following.


First, note that in two dimensions, the moment applied by member $i$ acting on one of its anchors at node $k$ is a scalar quantity:

\vspace{-0.2cm}
\[
M_i^k = -z_k F_i^x + x_k F_i^z.
\]

\noindent From the previous sections, $\bA \bq \in \mathbb{R}^{(nd)}$ are the forces applied by each member at each node, expressed component-wise in each direction.
So, by defining the moment arm matrix

\begin{equation}
    \bB = [-\bZ \; \; \bX],
\end{equation}

\noindent where $\bX$ and $\bZ$ are diag($\bx$) and diag($\bz$), the sum of the moments due to all members on each node is

\vspace{-0.1cm}
\[
\bM = \bB \bA \bq \quad \in \mathbb{R}^n.
\]

\noindent By removing the moment contributions at the stationary nodes, and removing the contributions of the bar members, the moments on the moving vertebra from the cables acting on its nodes are

\vspace{-0.1cm}
\begin{equation}
    \bM_c = \bW \bB \bA \bH \bq_s \quad \in \mathbb{R}^\eta.
\end{equation}

\noindent As with the nodal forces, the moments can be collapsed per body.
Since there is only one body remaining, the constraint matrix here becomes

\vspace{-0.1cm}
\begin{equation}
    \bA_m = \bOnes_\eta^\top \bW \bB \bA \bH.
\end{equation}

\noindent The moments from the external forces on the moving body are calculated in the same way,

\begin{equation}
    \bp_m = \bOnes_\eta^\top \bW \bB \bp,
\end{equation}

\noindent and so the moment balance for the moving vertebra is

\begin{equation}\label{eq:Am_pm}
    \bA_m \bq_s = \bp_m.
\end{equation}

\subsubsection{Combined static equilibrium constraint}

The force and moment balance conditions, eqns. (\ref{eq:Aqp_rigidbody}) and (\ref{eq:Am_pm}), can then be combined by stacking the systems of equations, as in

\begin{equation}\label{eq:Ab_pb_stacked}
    \bA_b = \begin{bmatrix} \bA_f \\ \bA_m \end{bmatrix}, \quad
    \bp_b = \begin{bmatrix} \bp_f \\ \bp_m \end{bmatrix},
\end{equation}


\noindent so that the full static equilibrium condition is

\begin{equation}\label{eq:Ab_pb}
    \bA_b \bq_s = \bp_b.
\end{equation}

\noindent Though the static equilibrium constraint has been fundamentally transformed from a per-node force balance into a per-body force and moment balance, the constraint is still linear.
This allows the application of the same approaches to solving eqn. (\ref{eq:Ab_pb}) as are done in prior literature for eqn. (\ref{eq:Aqp_forcedensity}).

For the spine robot in this research, eqn. (\ref{eq:Ab_pb}) has solutions.
The matrix $\bA_b \in \mathbb{R}^{6 \times 4}$ has rank 3 in all poses described by the specified $\bar \bxi$, thus a null space of dimension $4-3=1$.
Simulations below also show that eqn. (\ref{eq:Ab_pb}) is consistent.

\subsection{Inverse Statics Optimization}

With the static equilibrium condition in hand, an inverse statics optimization problem can be posed to find the optimal cable tensions that satisfy eqn. (\ref{eq:Ab_pb}).
The term `inverse statics' is used here to emphasize that a control system chooses a $\bq_s$.
For comparison, the forward statics problem would specify the cable model by fixing $\bar \bu$, and solving for the $\bq$ that evolves naturally due to the applied load $\bp$.
Here, instead, the trajectory generation problem solves for an optimal $\bq_s$ for a given $\bp$, and back-calculates the corresponding inputs $\bar \bu$.

Since eqn. (\ref{eq:Ab_pb}) is a linear system, and therefore a linear equality constraint, solving for an optimal $\bq_s^*$ can be made convex.
In particular, with a quadratic cost on $\bq_s$, the following quadratic program can be used:

\begin{align} 
\bq_s^* = \argmin_{\bq_s} \; \; & \bq_s^\top \bq_s \label{eq:forcedensityrigid_opt_obj}\\
\text{s.t.} \; \; & \bA_b \bq_s = \bp_b \label{eq:forcedensityrigid_opt_Aqp}\\
                  & -\bq_s \leq - \bc. \label{eq:forcedensityrigid_opt_mintension}
\end{align}

\noindent The inequality constraint (\ref{eq:forcedensityrigid_opt_mintension}) is used here to enforce positive cable tensions, where $\bc$ is a constant vector specifying the minimum force density in each cable.
Similar optimization problems have been termed ``inverse kinematics'' in the literature \cite{friesen2014,Friesen2016}, and this work is differentiated by the fundamental modification of eqn. (\ref{eq:forcedensityrigid_opt_Aqp}).

The inverse statics procedure first calculates the constraint eqn. (\ref{eq:forcedensityrigid_opt_Aqp}) for each $\bar \bxi(t)$ in the reference state trajectory, via eqns. (\ref{eq:A_forcedensity})-(\ref{eq:p_forcedensity}) and (\ref{eqn:W_defn})-(\ref{eq:Ab_pb}).
Then (\ref{eq:forcedensityrigid_opt_obj}-\ref{eq:forcedensityrigid_opt_mintension}) is solved for each $\bq_s^*(t)$ via an optimization solver, and $\bar \bu(t)$ is calculated using eqn. (\ref{eqn:backcalculate_u}).
The procedure follows Algorithm (\ref{alg:invstat}).

\vspace{0.2cm}
\begin{algorithm}[H]
    \caption{Reference Input Trajectory Generation}\label{alg:invstat}
    \begin{algorithmic}
        \Procedure{InvStat}{$\bar \bxi$}
        {\fontsize{10pt}{14pt}\selectfont
        \For{$t \gets 1, T$}
            \State $\bA_b \gets \bA_b(\bx(\bar \bxi(t)), \; \bz(\bar \bxi(t)))$
            \State $\bp_b \gets \bp_b(\bx(\bar \bxi(t)), \; \bz(\bar \bxi(t)))$
            \State $\bq_s^*(t) \gets OptForceDens(\bA_b, \; \bp_b)$ \Comment{solve (\ref{eq:forcedensityrigid_opt_obj}-\ref{eq:forcedensityrigid_opt_mintension})}
            \For{$i \gets 1, s$}
                \State $\bar u_i(t) \gets \ell_i(t) - \dfrac{\ell_i(t) q_i^*(t)}{k_i}$
            \EndFor
        \EndFor
        } 
        \State \textbf{return} $\bar \bu$
        \EndProcedure
    \end{algorithmic}
\end{algorithm}
\vspace{0.1cm}

\noindent Algorithm (\ref{alg:invstat}) was implemented with MATLAB's {\tt{quadprog}} solver in the software supplied by the authors\footnote{https://github.com/BerkeleyExpertSystemTechnologiesLab/ultra-spine-simulations}.

\section{Controller Formulations}\label{sec:controller}

\begin{figure}[htpb]
    \centering
    \begin{subfigure}{1.0\columnwidth}
        \centering
        \includegraphics[width=0.8\columnwidth]{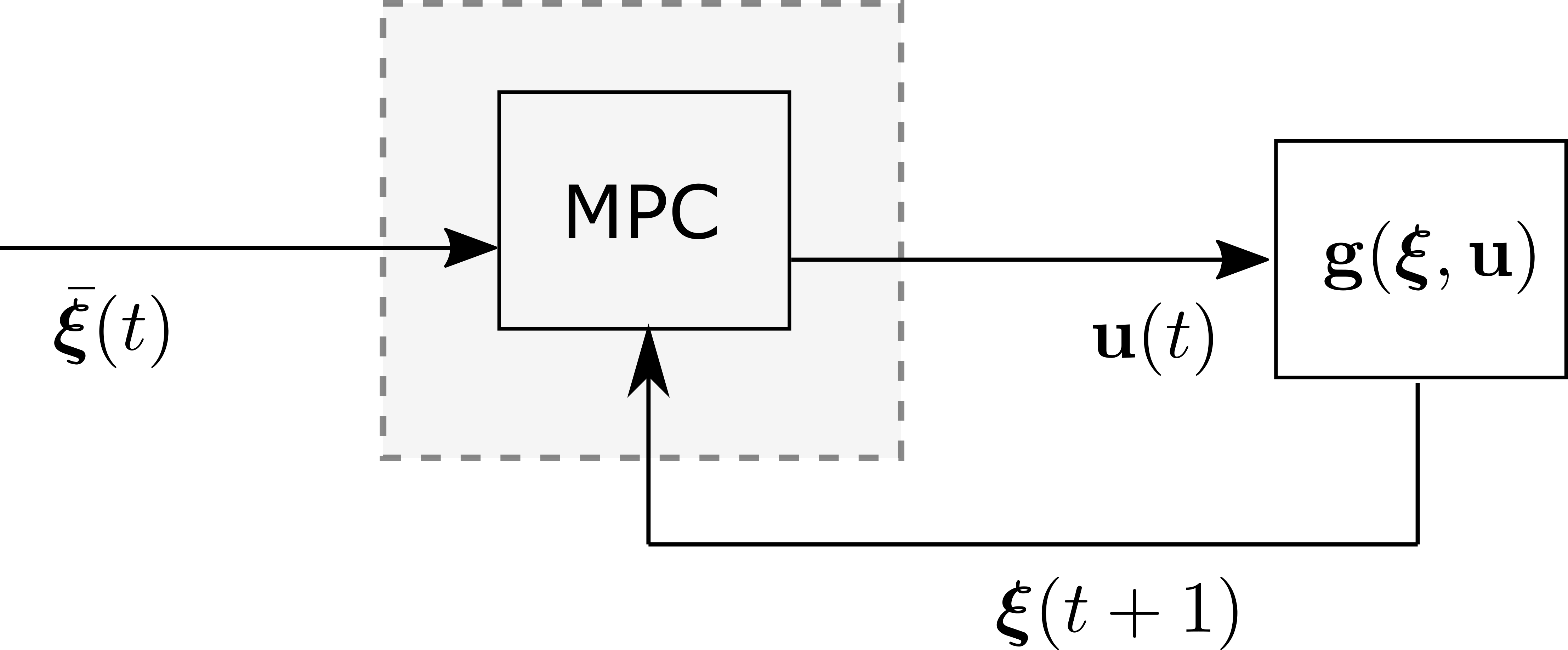}
        \caption{Block diagram for the model-predictive controller with smoothing terms, no input trajectory generation.}
        \label{fig:MPC_block_diagram}
        \vspace{0.3cm}
    \end{subfigure}
    \begin{subfigure}{1.0\columnwidth}
        \centering
        \includegraphics[width=0.8\columnwidth]{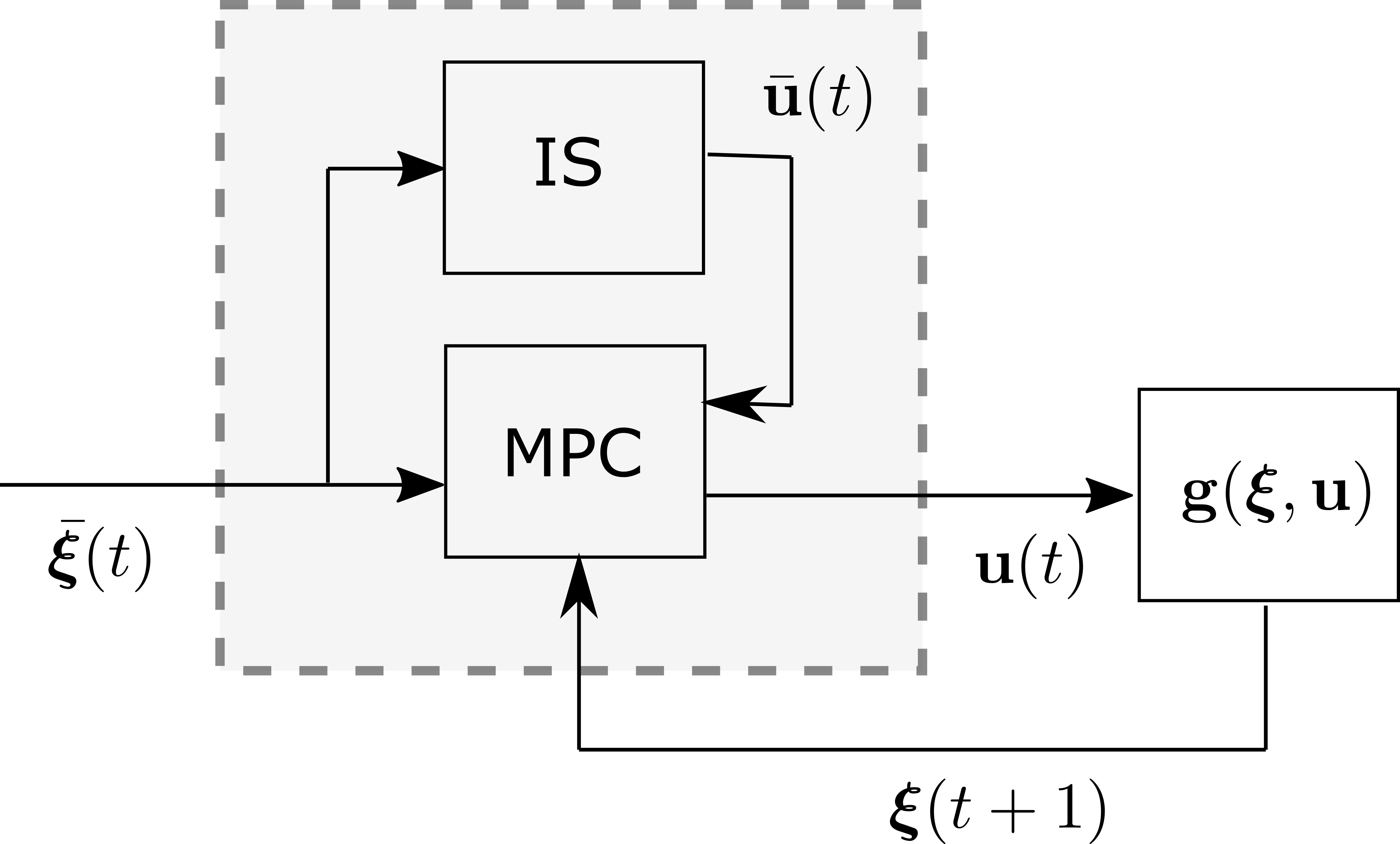}
        \caption{Block diagram of the proposed controller that combines inverse statics (IS) for input trajectory generation with model-predictive control (MPC) to close the loop.}
        \label{fig:MPC_IS_block_diagram}
    \end{subfigure}
    \vspace{0.2cm}
    \caption{Block diagrams of the two controllers considered in this work. Both controllers are simulated against the ground-truth nonlinear dynamics $\bg(\bxi, \bu)$.}
    \label{fig:block_diagrams}
    \vspace{-0.1cm}
\end{figure}

This work introduces two controllers for the tensegrity spine robots under consideration.
The first uses a model-predictive control (MPC) law, and incorporates smoothing terms into the optimization problem (Fig. \ref{fig:MPC_block_diagram}.)
The second uses the inverse statics routine, Algorithm (\ref{alg:invstat}), for reference input trajectory generation, and a simplified version of the model-predictive control law to close the control loop (Fig. \ref{fig:MPC_IS_block_diagram}.) 
Both controllers incorporate a linearization of the equations of motion, eqn. (\ref{eq:ss_definition}), in the control calculations; however, all are simulated against the ground-truth nonlinear system.

As discussed in Sec. \ref{sec:background_mpc}, this work prioritizes practicality over theoretical guarantees.
For this reason, neither formulation contains terminal constraints, and thus stability can only be shown experimentally, not proven.

The following sections use subscripts (e.g., $\bu_t$) to represent predicted values of vectors at a time instance, and parentheses (e.g., $\bu(t)$) to represent a measured or applied value at that time instance.
Note that these are the same for the reference trajectory ($\bar \bu(t) = \bar \bu_t$) and so the notation is used interchangeably.
Superscripts (e.g., $\bxi^{(i)}$) index into a vector.

\vspace{-0.1cm}
\subsection{Model-Predictive Controller Formulation}

For both controllers, the MPC block generates a control input $\bu(t)$ via the following.
At each timestep $t$, a constrained finite-time optimal control problem (CFTOC) is solved, generating the sequence of optimal control inputs $\bU^*_{t\rightarrow t+N|t} = \{{\bu}^*_{t|t}, ..., {\bu}^*_{t+N|t}\}$, over a horizon of $N$ timesteps ahead.
The notation $t+k|t$ represents a value at the timestep $t+k$, as predicted at timestep $t$ (from \cite{Borrelli2003}, Ch. 4.)
The first input ${\bu}^*_{t|t}$ is applied, as in $\bu(t) = \bu^*_{t|t}$, closing the loop.
The following sections define this CFTOC problem for each case, fully specifying the controllers.

\subsection{Controller with MPC and Smoothing Terms}\label{sec:ctler_smoothing}

The first controller (presented in the conference version of this work, \cite{Sabelhaus2017}) adapts the standard linear time-varying MPC formulation by adding a variety of hand-tuned weights and constraints. This is a common approach to establishing proof-of-concept control \cite{Falcone2007}, which was the goal in \cite{Sabelhaus2017}. \\

\subsubsection{Constrained Finite-Time Optimal Control Problem Formulation}

The following CFTOC problem is solved at each timestep $t$ using a quadratic programming solver.
Here, $N=10$ is the horizon length and $w_1...w_7$ are constant scalar weights.
The functions $p$ and $q$ represent the terminal cost and stage cost of the objective function, not to be confused with the inverse statics force balance of Sec. \ref{sec:invstat}.
The objective function, and the use and purpose of the constraints, are given in subsections \ref{sec:smoothing_dynamics_constraint} to \ref{sec:smoothing_obj_fcn}.

\vspace{-0.3cm}
\begin{align}
\displaystyle\min_{\bU_{t\rightarrow t+N|t}} \; \; & p(\bxi_{t+N|t}, \Delta \bxi_{t+N|t}) \quad \dots \notag \\
& + \sum_{k = 0}^{N-1} q(\bxi_{t+k|t}, \Delta \bxi_{t+k|t}, \Delta \bu_{t+k|t} ) \label{eq:opt_main} \\
\text{s.t.} \; \; & \bxi_{t+k+1|t} = \bA_{t} \bxi_{t+k|t} + \bB_{t} \bu_{t+k|t} +\bc_{t}  \label{eq:sys_dynamics} \\
& \Delta \bxi_{t+k|t} = \bxi_{t+k|t} - \bxi_{t+k-1|t}  \label{eq:delta_xi} \\
& \Delta \bu_{t+k|t} = \bu_{t+k|t} - \bu_{t+k-1|t}  \label{eq:delta_u} \\
& \bxi_{t|t} = \bxi(t) \label{eq:assign_xi} \\
& \bu^{min} \leq \bu_{t+k} \leq \bu^{max} \label{eq:u_lim} \\
& \| \bu_{t|t} - \bu_{t-1} \|_\infty \leq w_1  \label{eq:smoothing_u1} \\
& \| \bu_{t+k|t} - \bu_{t|t} \|_\infty \leq w_2, \; k=1..(N-1) \label{eq:smoothing_uk} \\
& \| \bu_{t+N|t} - \bu_{t|t} \|_\infty \leq w_3 \label{eq:smoothing_uN} \\
& \| \Delta \bxi^{(1:6)}_{t+k|t} \|_\infty \leq w_4 \label{eq:smoothing_x1} \\
& \| \Delta \bxi^{(13:18)}_{t+k|t}  \|_\infty \leq w_5 \label{eq:smoothing_x2} \\
& \| \Delta \bxi^{(25:30)}_{t+k|t} \|_\infty \leq w_6 \label{eq:smoothing_x3} \\
& \bxi^{(3)}_{t+k|t} + w_7 \leq \bxi^{(15)}_{t+k|t} \label{eq:collision_2} \\
& \bxi^{(15)}_{t+k|t} + w_7 \leq \bxi^{(27)}_{t+k|t}. \label{eq:collision_3}
\end{align}

\vspace{0.2cm}

\subsubsection{Dynamics Constraint}\label{sec:smoothing_dynamics_constraint}

The dynamics constraint (\ref{eq:sys_dynamics}) consists of a time-varying linearization of the system, as in:


\begin{align}
& \bA_t = \frac{\partial \bg(\bxi, \bu)}{\partial \bxi} \Bigr|_{ \substack{ \bxi = \bxi_{t-1} \\ \bu = \bu_{t-1} }} \label{eq:linearization_A}\\
& \bB_t = \frac{\partial \bg(\bxi, \bu)}{\partial \bu} \Bigr|_{ \substack{ \bxi = \bxi_{t-1} \\ \bu = \bu_{t-1} }}  \label{eq:linearization_B}
\\
& \bc_t = \bg(\bxi_{t-1}, \bu_{t-1}) - \bA_t \bxi_{t-1} - \bB_t \bu_{t-1}. \label{eq:linearization_c}
\end{align}

This linearization (\ref{eq:linearization_A}-\ref{eq:linearization_c}) is implemented as a finite difference approximation.
This approach is chosen due to computational issues with calculating additional analytical derivatives of the dynamics.
The calculated $\bA_t, \bB_t, \bc_t$ are used over the entire horizon.
For the start of the simulation, $\bu_{0} = \bzero$ is used.
Since these linearizations are not at equilibrium points, the linear system is affine, with $\bc_t$ being a constant vector offset. 

Here, the continuous-time linearized dynamics are used as a constraint, and are not discretized.
Since the timesteps in the simulations below are small ($dt=0.001$ sec.), a discretization does not significantly alter the values of $\bA_t, \bB_t, \bc_t$.\\


\subsubsection{Other Constraints}

The remaining constraints are either smoothing terms, constraints motivated by the physical robot, or miscellaneous housekeeping terms.

Constraints (\ref{eq:delta_xi}) and (\ref{eq:delta_u}) define the $\Delta \bu$ and $\Delta \bxi$ variables, which are used for the smoothing constraints on the inputs and states.
Constraint (\ref{eq:assign_xi}) assigns the state variable at the start of the optimization horizon, $\bxi_{t|t}$, to the actual observed value of the state from the previous simulation timestep, $\bxi(t)$.

Constraint (\ref{eq:u_lim}) is a bound on the inputs, limiting the length of the cable rest lengths, with $\bu^{min},\bu^{max} \in \mathbb{R}^{24}$ but having the same value for all inputs (Table \ref{tab:weights}).

Constraints (\ref{eq:smoothing_u1}-\ref{eq:smoothing_x3}) are smoothing terms to compensate for the lack of an reference input trajectory.
Of these, (\ref{eq:smoothing_u1}-\ref{eq:smoothing_uN}) are for the inputs, where $\bu_{t-1}$ is the most recent input at the start of the CFTOC problem.
Constraints (\ref{eq:smoothing_x1}-\ref{eq:smoothing_x3}) are smoothing terms on the states, limiting the deviation between successive states in the trajectory.
These reduce linearization error, and are split so that the positions and angles of each vertebra could be weighted differently.
No velocity terms are constrained.

Finally, since states $\{ \bxi^{(3)}, \bxi^{(15)}, \bxi^{(27)}
\}$ are the vertebra $z$-positions, constraints (\ref{eq:collision_2}-\ref{eq:collision_3}) prevent vertebra collisions. \\

\subsubsection{Objective Function}\label{sec:smoothing_obj_fcn}

The objective function has a terminal cost $p$ and a stage cost $q$ defined as the following.
Here, shortened notation such as $\lVert \Delta \bxi_{t+k|t} \rVert_{\bS^{k}}^2$ denotes a weighted quadratic term, as in $(\Delta \bxi_{t+k|t} )^\top \bS^k (\Delta \bxi_{t+k|t}).$

\begin{align}
p( \bxi_{t+N|t}, \Delta \bxi_{t+N|t} )  = & \; \lVert \bxi_{t+N|t} - \bar \bxi_{t+N|t} \rVert_{\bQ^{N}}^2 \notag \\ 
& + \lVert \Delta \bxi_{t+N|t} \rVert_{\bS^{N}}^2, \label{eq:termcost}\\[0.3cm]
q(\bxi_{t+k|t}, \Delta \bxi_{t+k|t}, \Delta \bu_{t+k|t} ) = & \; \lVert \bxi_{t+k|t} - \bar \bxi_{t+k|t} \rVert_{\bQ^{k}}^2 \notag \\
& \; + \lVert \Delta \bxi_{t+k|t} \rVert_{\bS^{k}}^2 \notag \\
& \; + w_8  \lVert \Delta \bu_{t+k|t} \rVert_{\infty}. \label{eq:stagecost}
\end{align}

\vspace{0.2cm}

As before, $w_8$ is a scalar, while $\bQ$ and $\bS$ are constant diagonal weighting matrices which are exponentiated by the timestep in the optimization horizon.
Here, $\bQ$ penalizes the tracking error in the states, $\bS$ penalizes the deviation in the states at one timestep to the next, and $w_8$ penalizes the deviations in the inputs from one timestep to the next.
These matrices are diagonal, with blocks corresponding to the Cartesian and Euler angle coordinates, with zeros for all velocity states, according to vertebra.

Raising each diagonal element of $\bQ$ or $\bS$ to the power of $k$ or $N$ puts a heavier penalty on terms farther away on the horizon.
These are defined as:

\vspace{-0.1cm}
\begin{align}\label{eq:QR}
& \bar \bQ^k = diag( w_9^k, \: w_9^k, \: w_9^k \: | \: w_{10}^k, \: w_{10}^k, \: w_{10}^k \: | \: 0 ... 0) \in \mathbb{R}^{12 \times 12} \notag\\
& \bar \bS^k = diag( w_{11}^k, \: w_{11}^k, \: w_{11}^k \: | \: w_{11}^k, \: w_{11}^k, \: w_{11}^k \: | 0...0) \in \mathbb{R}^{12 \times 12} \notag\\
& \bQ^k = \bI_3 \otimes \bar \bQ^k, \quad \quad \bS^k = \bI_3 \otimes \bar \bS^k.
\end{align}



\noindent Table \ref{tab:weights} lists all the constants for this controller, including the constraints and the objective function, with units.

\begin{table}[!ht]
\centering
\caption{Smoothing controller weights and constants.}
\label{tab:weights}
\begin{tabularx}{1\columnwidth}{| l | X  l | l |}
    \hline 
    Constant: & Value: & & Interpretation:\\
    \hline
    $N$       & 10  & no units & Horizon Length \\
    $u_{min}$ & 0.0 &  meters (cable) & Min. Cable Length \\
    $u_{max}$ & 0.20  &  meters (cable) & Max. Cable Length \\
    $w_1$     & 0.01 &  meters (cable) & Input Smooth., Horiz. Start \\
    $w_2$     & 0.01 &  meters (cable) & Input Smooth., Horiz. Middle \\
    $w_3$     & 0.10 &  meters (cable) & Input Smooth., Horiz. End \\
    $w_4$     & 0.02 & meters and radians & State Smooth., Bottom Vert. \\
    $w_5$     & 0.03 & meters and radians & State Smooth., Mid. Vert. \\
    $w_6$     & 0.04 & meters and radians & State Smooth., Top Vertebra \\
    $w_7$     & 0.02 & meters & Vertebra Anti-Collision \\
    $w_8$     & 1   & no units & Input Smoothing \\
    $w_9$     & 25   & no units & State Tracking, Vertebra Pos. \\
    $w_{10}$  & 30    & no units & State Tracking, Vert. Angle \\
    $w_{11}$  & 3    & no units & Input Difference Penalty \\
    \hline 
\end{tabularx}
\end{table}

\subsection{Controller with MPC and Inverse Statics Optimization}\label{sec:ctlr_invstat}




A primary contribution of this work (in comparison to \cite{Sabelhaus2017}) is a controller that combines the inverse statics (IS) optimization, via Algorithm (\ref{alg:invstat}), with an MPC block.
As shown in Fig. \ref{fig:MPC_IS_block_diagram}, the IS block generates a reference input trajectory $\bar \bu$ that is tracked alongside $\bar \bxi$ as part of the MPC problem.
This approach contributes a new method to address computational complexity and tuning.
The IS solutions can be solved offline, reducing the load on the MPC optimization problem.
This approach also significantly reduces hand-tuning.
As discussed in this section, the controller is formulated for the 2D, single-vertebra spine model. \\


\subsubsection{Constrained Finite-Time Optimal Control Problem Formulation}

The following CFTOC problem is solved at each timestep $t$ using a quadratic programming solver. 


\begin{align}
\displaystyle\min_{\bU_{t \rightarrow t+N|t}}
\quad & p(\bxi_{t+N|t}) + \sum_{k = 0}^{N-1} q(\bxi_{t+k|t}, \bu_{t+k|t}) \label{eq:opt_main_2d}\\
\text{s.t.} \quad & \bxi_{t+k+1|t} = \bA_t \bxi_{t+k|t} + \bB_t \bu_{t+k|t} + \bc_t \label{eq:sys_dynamics_2d} \\
    & \bxi_{t|t} = \bxi(t) \label{eq:initialcondition_2d} \\
    & \bu_{t+k|t} \geq \bu^{min} \label{eq:u_lim_2d}\\
    & \bxi^{(2)}_{t+k|t} \geq w_1. \label{eq:collision_2d}
\end{align}
\vspace{0.1cm}

This formulation (\ref{eq:opt_main_2d}-\ref{eq:collision_2d}) is significantly simpler than the smoothing formulation (\ref{eq:opt_main}-\ref{eq:collision_3}), with only one scalar tuning weight $w_1$, and a much smaller horizon length (Table \ref{tab:weights_invstat}).
As above, $p$ and $q$ represent the terminal cost and stage cost of the objective function.
The following sections define the objective function, and use and purposes of the constraints. \\


\subsubsection{Dynamics Constraint}

Constraint (\ref{eq:sys_dynamics_2d}) enforces the time-varying linearized system dynamics, just as with the smoothing controller, via eqns. (\ref{eq:linearization_A}-\ref{eq:linearization_c}).
The two-dimensional controller also applies a zero-order hold to (\ref{eq:sys_dynamics_2d}) for increased prediction fidelity.
However, due to the small timesteps involved, the values of $\bA_t$, $\bB_t$, and $\bc_t$ remained mostly unchanged after this operation, with no noticeable effect on simulation results.\\

\subsubsection{Other Constraints}

The remaining constraints have the same interpretations as their counterparts in the smoothing controller formulation.
Constraint (\ref{eq:initialcondition_2d}) assigns the initial condition at the starting time of the CFTOC problem.
Constraint (\ref{eq:u_lim_2d}) is a linear constraint on the inputs so that the cables cannot have negative rest lengths.
Finally, constraint (\ref{eq:collision_2d}) denotes a minimum bound on the second element in the state, the $z$-position, which prevents collision between the moving vertebra and the static vertebra. \\

\subsubsection{Objective Function}

The objective function for this formulation is comprised of quadratic weights on the state and input tracking errors.
As opposed to the smoothing formulation, which included non-traditional terms, the objective function here is exactly the same as with standard MPC.
Using similar notation as in equations (\ref{eq:termcost}) and (\ref{eq:stagecost}),

\begin{align}
p(\bxi_{t+N|t}) = & \lVert \bxi_{t+N|t} - \bar \bxi_{t+N|t} \rVert^2_{\bQ}, \label{eq:terminalcost_invstat}\\
q( \bxi_{t+k|t}, \bu_{t+k|t}) = & \lVert \bxi_{t+k|t} - \bar \bxi_{t+k|t} \rVert^2_{\bQ} \notag\\
& + \lVert \bu_{t+k|t} - \bar \bu_{t+k|t} \rVert^2_{\bR}. \label{eq:stagecost_invstat}
\end{align}

Here, $\bQ$ and $\bR$ are constant diagonal weighing matrices which penalize state and input tracking errors respectively, defined similarly to the smoothing formulation, but do not vary with the horizon step as with the $\bQ^k$ terms in eqn. (\ref{eq:stagecost}).
Specifically, these weights are

\begin{align}
& \bQ = diag( w_2, \: w_2, \: w_2 \: | \: 0 ... 0) \in \mathbb{R}^{6 \times 6} \label{eq:Q_invstat},\\
& \bR = diag( w_3, \: w_3, \: w_3, \: w_3) \in \mathbb{R}^{4 \times 4}.  \label{eq:R_invstat}
\end{align}

As with eqn. (\ref{eq:QR}), the $\bQ$ matrix does not penalize velocity states.
Table \ref{tab:weights_invstat} lists all the constants for this controller, including the constraints and the objective function, with units.

\renewcommand{\arraystretch}{1.1}
\begin{table}[!ht]
\centering
\caption{Input tracking controller weights and constants.}
\label{tab:weights_invstat}
\begin{tabularx}{1\columnwidth}{| l | X  l | l |}
    \hline 
    Constant: & Value: & & Interpretation:\\
    \hline
    $N$       & 4   & no units & Horizon Length \\
    $u_{min}$ & 0.0 &  meters (cable) & Min. Cable Length \\
    $w_1$     & 0.075 &  meters (vertebra position) & Vertebra Anti-Collision \\
    $w_2$     & 1 &  no units & State Tracking Penalty \\
    $w_3$     & 10 & no units & Input Tracking Penalty \\
    \hline 
\end{tabularx}
\end{table}

\vspace{-0.1cm}
\subsection{Controller Comparison}\label{sec:controller_comparison}

The differences between the two controller formulations (Sec. \ref{sec:ctler_smoothing} and \ref{sec:ctlr_invstat}) are summarized in Table \ref{tab:comparison}.
In addition to the inherent difference between the tracking of one vertebra versus 3 vertebrae, and the difference between the 2D and 3D models that are tracked, three major considerations are present.

First, the controller with the IS optimization is much more general, and does away with the smoothing terms.
This reduces the complexity of the CFTOC problem, thus removing most of the need for tuning optimization weights (compare Table \ref{tab:weights} versus Table \ref{tab:weights_invstat}).
Second, the controller with the IS optimization moves some computational load offline, since the MPC problem now has fewer terms.
Third, in contrast to those benefits, the MPC plus IS controller required a faster simulation rate as tested here, with the discretization timestep of $dt=1e^{-5}$ versus $1e^{-3}$ for the smoothing controller.
These three changes represent the tradeoffs between tuning requirements and performance implications of either controller.

\section{Simulation Results}\label{sec:results}

Two sets of simulations are presented in this work, one for the controller with MPC and smoothing terms, and one for the controller with MPC and inverse statics reference input generation/tracking.
All simulation work used the YALMIP toolbox in MATLAB \cite{Lofberg2004}, with Gurobi as the solver.
All code is available online\footnote{https://github.com/BerkeleyExpertSystemTechnologiesLab/ultra-spine-simulations}. 



For both models and controllers, simulations are also performed with noise, in order to test closed-loop performance.
Appendix Sec. \ref{appendix:noise_model} gives the noise model in detail.

\subsection{Computational Performance}

The optimization problem for the MPC plus smoothing controller, applied to the 3D model (from Sec. \ref{sec:ctler_smoothing}) took $0.5-1$ sec. to solve at each timestep, using the Gurobi solver.
The optimization problem for the MPC plus inverse statics reference input tracking controller, applied to the 2D model (from sec. \ref{sec:ctlr_invstat}),  took $0.15-0.2$ sec. to solve at each timestep.
The inverse statics optimization procedure (Alg. \ref{alg:invstat}) is performed offline before the closed-loop tests begin, so is not timed; however, it solves rapidly enough for practical use.


\subsection{Controller with MPC and Smoothing Terms}\label{sec:results_smoothing}

Fig. \ref{fig:allvert} shows the paths of the vertebrae in the 3D, three-vertebra simulation, using the smoothing constraint controller, in the $X$-$Z$ plane as they sweep through their counterclockwise bending motion.
Fig. \ref{fig:allvert} includes the reference trajectory (blue), the resulting trajectory with the smoothing MPC controller and no noise (green), and a representative result of controller with added noise (magenta).
Fig. \ref{fig:topvert} shows a larger view of the top vertebra center of mass, which had the largest tracking errors of the three vertebrae, and which is used for comparison with the 2D single-vertebra model below.

The tracking errors for each state, for each vertebra, for both simulations (with and without noise) are shown in Fig. \ref{fig:all_errors}.
In both simulations, an initial transient is observed in the $X$-position and $\gamma$-angle states.
This is possibly due to a zero initial velocity of the vertebrae, requiring the spine to rapidly move at the start of its simulation to ``catch up" with the trajectory.
After that, all errors trend to zero, with the expected oscillations in the simulation with noise.

\subsection{Controller with MPC and Inverse Statics Optimization}\label{sec:results_invstat}

\begin{figure}[thpb]
    \centering
    \includegraphics[width=1\columnwidth]{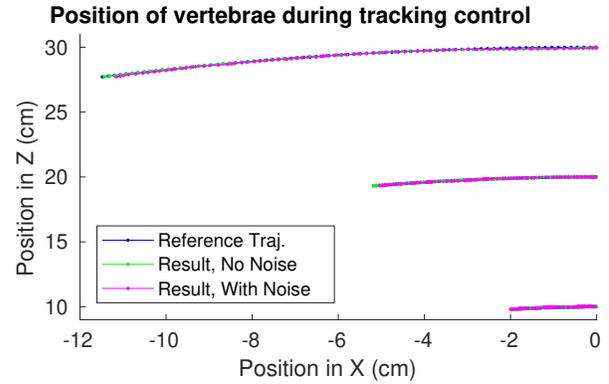}
    \caption{Positions in the $X$-$Z$ plane of all vertebrae, for the 3D, three-vertebra model with the smoothing controller. Plot includes the state reference and the two simulations (with/without noise), as the robot performs the counterclockwise bend described in Fig. \ref{fig:spine_traj}.}
    \label{fig:allvert}
\end{figure}

\begin{figure}[thpb]
    \centering
    \includegraphics[width=0.95\columnwidth]{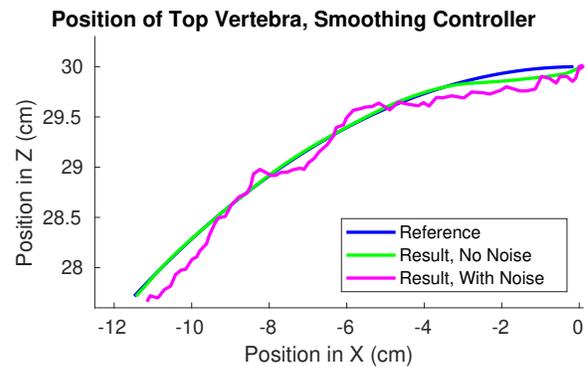}
    \caption{Positions in the X-Z plane of the top vertebra, for the 3D, three-vertebra model with the smoothing controller, including the state reference and the two simulations (with/without noise), as the robot performs a counterclockwise bend. The vertebra tracks the trajectory with small error.}
    \label{fig:topvert}
\end{figure}

\begin{figure}[thpb]
    \centering
    \includegraphics[width=0.95\columnwidth]{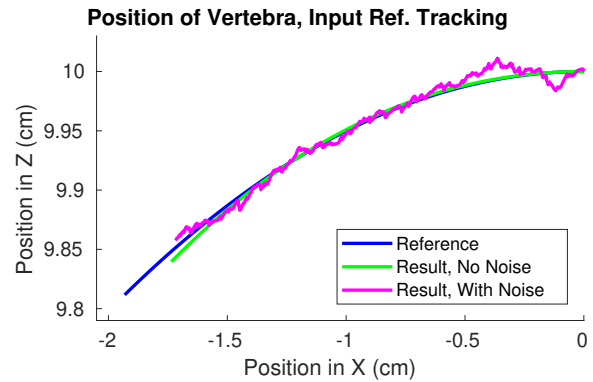}
    \caption{Positions in the X-Z plane of the single vertebra for the 2D model, using the controller with MPC plus inverse statics reference input trajectory generation/tracking, including the state reference and the two simulations (with/without noise), as the robot performs a counterclockwise bend. The vertebra tracks the trajectory, but accumulates more lag in comparison to the smoothing controller (Fig. \ref{fig:topvert}.)}
    \label{fig:2d_vert_tracking}
\end{figure}

Fig. \ref{fig:2d_vert_tracking} shows the path of the single vertebra in the 2D simulation, using the controller with MPC plus inverse statics reference input tracking, as it sweeps through its counterclockwise bending motion.
As with Fig. \ref{fig:allvert} and \ref{fig:topvert}, the reference state trajectory is included (in blue) alongside results from the controller with no noise (green) and from a representative simulation with noise (magenta.) 
The vertebra follows the path of the of the kinematic states, but experiences some accumulation of lag.
The results show that the closed-loop controller is noise-insensitive, alongside accurate tracking, but that the lag occurs in all circumstances.

\begin{figure*}[tb]
    \centering
    \begin{subfigure}[h]{1\textwidth}
        \centering
        \includegraphics[width=0.9\textwidth]{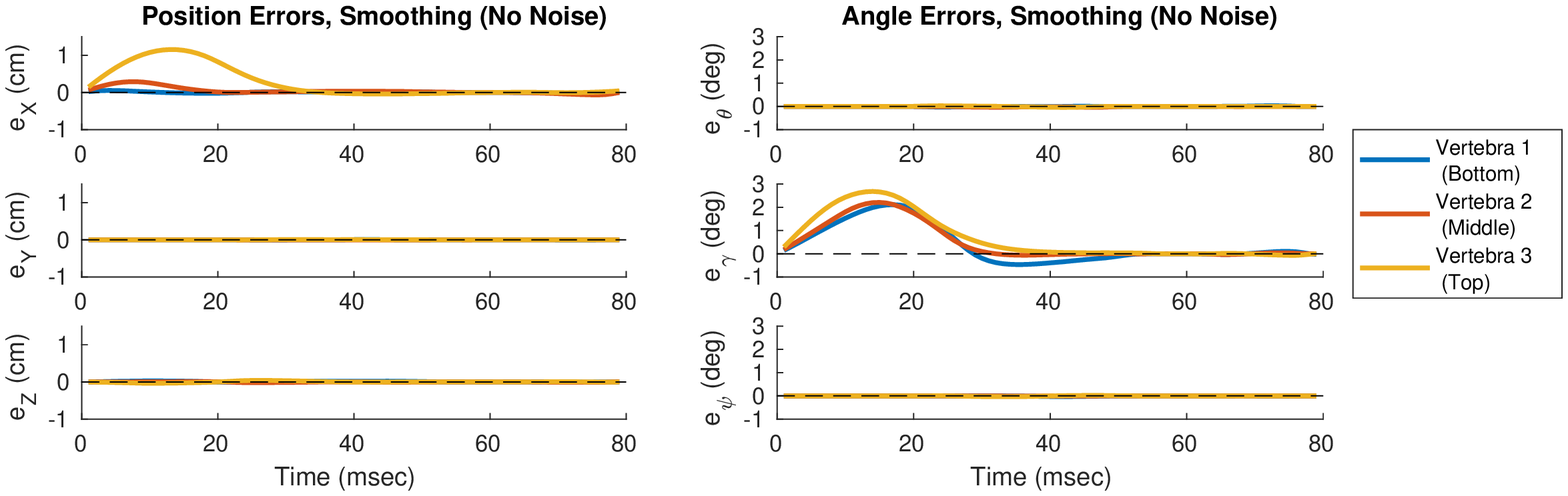}
    \end{subfigure}
    \begin{subfigure}[h]{1\textwidth}
        \centering
        \includegraphics[width=0.9\textwidth]{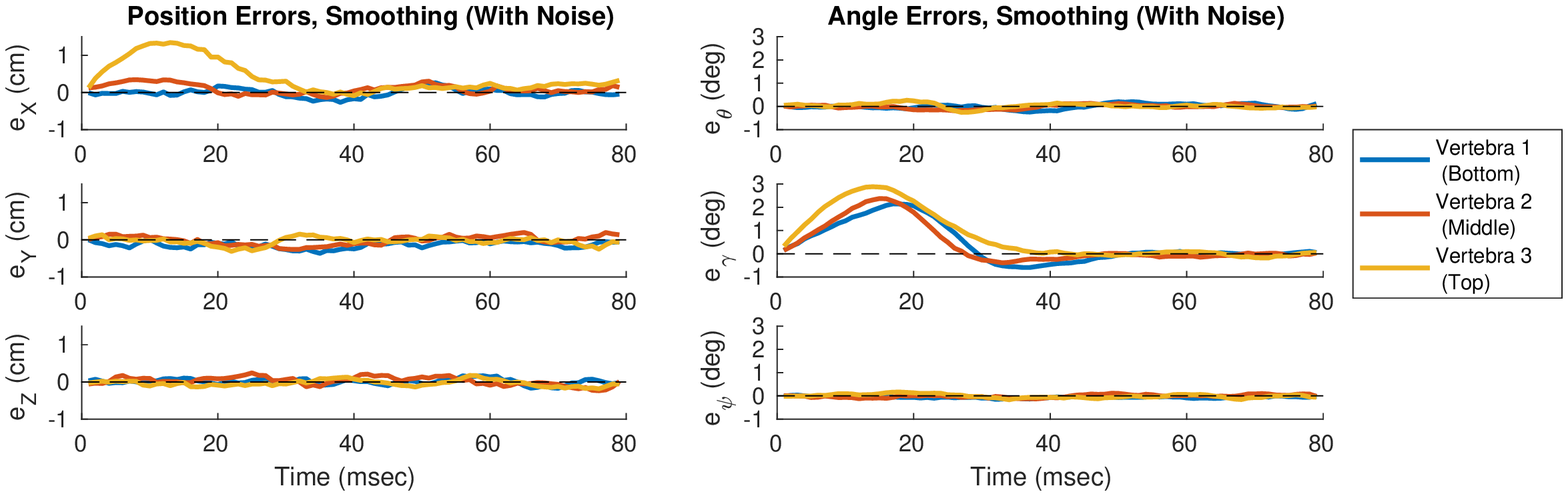}
    \end{subfigure}
    \vspace{-0.2cm}
    \caption{Tracking errors in system states for the 3D, three-vertebra model using the smoothing controller, with and without noise. Position states ($x,y,z$) on the left with units of cm, Euler angles ($\theta,\gamma,\psi$) on the right with units of degrees.}
    \label{fig:all_errors}
    \vspace{-0.1cm}
\end{figure*}

\begin{figure}[thpb]
    \centering
    \includegraphics[width=1\columnwidth]{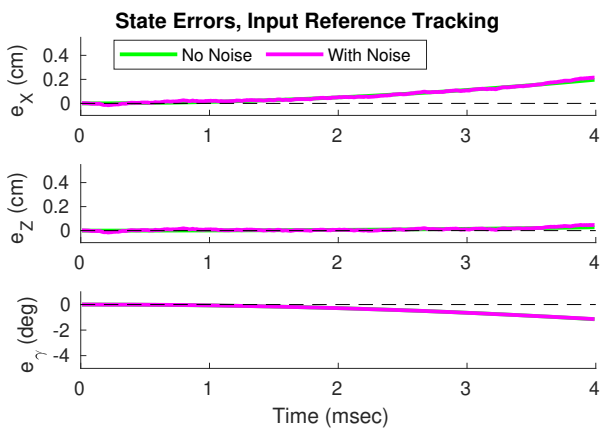}
    \caption{Tracking errors in system states for the 2D, single-vertebra model using the controller with MPC plus inverse statics reference input trajectory generation/tracking, with and without noise. Position errors are in cm, rotation errors are in degrees. The drift shown here arises from controller lag.}
    \label{fig:all_errors_2d}
    \vspace{-0.2cm}
\end{figure}

The tracking errors for each state are shown in Fig. \ref{fig:all_errors_2d}, using the same convention as Fig. \ref{fig:all_errors}.
The controller accumulates lag throughout the simulation, and the errors do not converge.
This is expected, since the tracked inputs are inverse statics and not dynamics, and this simulation setup violates the assumption of quasi-static movement.
Since the results presented here are used to compare with the smoothing controller, the simulations use the same setup with only dynamic movement.
It is expected that given a setup where the controller has the opportunity to settle, the errors would converge.

\section{Control of Different Spines}\label{sec:different_spines}

\begin{figure}[thpb]
    \vspace{-0.1cm}
    \centering
    \includegraphics[width=0.95\columnwidth]{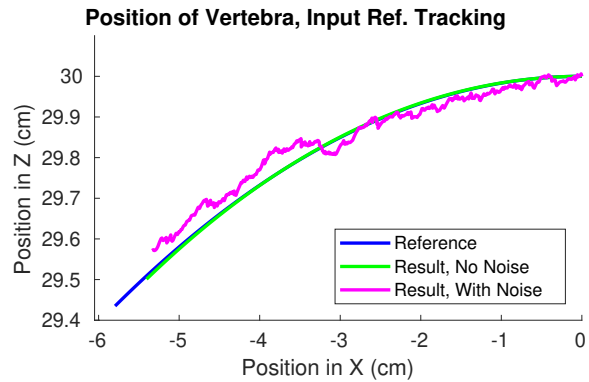}
    \caption{Additional test of MPC plus IS controller with the larger, differently-shaped vertebra. This controller tracks this vertebra in the same way as Fig. \ref{fig:2d_vert_tracking} with no need to change any tuning constants.}
    \label{fig:2d_vert_tracking_larger}
    \vspace{-0.3cm}
\end{figure}

The proposed controller that combines MPC with inverse statics for reference input tracking has significantly fewer tuning parameters.
It is thus easily extendable to different sizes and shapes of spines, whereas a large amount of tuning may have otherwise been required.

In order to illustrate this, the controller was tested on a different 2D spine, with a different size and shape of vertebra.
Control results (Fig. \ref{fig:2d_vert_tracking_larger}) show equivalent performance to the original vertebrae of Sec. \ref{sec:results}, despite the size and geometry change.
For these tests, no changes were made to the inverse statics algorithm, nor to any of the constants in Table \ref{tab:weights_invstat}.




This differently-shaped spine still retained the same number of point masses (to satisfy the assumptions of the inverse statics algorithm), but is now larger and heavier, with different angles between its bars. 
These changes are motivated by ongoing designs of hardware prototypes.
The geometry, constants, and simulation setup are discussed in Appendix Sec. \ref{appendix:eqns_of_motion}.

\section{Discussion}\label{sec:discussion}

Both controllers exhibit state tracking characteristics which could be used in different environments for effective closed-loop control.
The smoothing controller tracked with lower error, after an initial transient response, but had higher computational complexity and tuning requirements.
The controller with inverse statics tracking, which is more general, exhibited lag and thus larger tracking errors, but with lower computational overhead and with significantly less hand-tuning. 
This is the first work (with \cite{Sabelhaus2017}) that tracks a state-space trajectory of a tensegrity spine robot in closed-loop, and the first which shows noise insensitivity.


\subsection{Computational Performance}

The lengths of time taken to solve the optimization problem for each controller ($0.5$-$1$ sec. and $0.15$-$0.2$ sec.) were longer than the timesteps of each respective simulation ($1e^{-3}$ and $1e^{-5}$ sec.).
Thus, the optimization procedure will need to be made more efficient before using this controller in hardware.
One approach that may reduce solver time is the calculation of a symbolic Jacobian for the $\bA_t$ and $\bB_t$ matrices, reducing the computational load in the linearization.


\subsection{Tracking Performance Comparison}

The controller with MPC and the inverse statics optimization removed the need for hand-tuned smoothing terms, but exhibited lag in tracking a highly-dynamic state trajectory.
This motivates the use of either controller in different settings.
The MPC plus smoothing controller may be appropriate for high-performance dynamic tracking, when the control system designer is able to tune the weights and constraints.
In contrast, the MPC plus inverse statics optimization controller may be appropriate for more pseudo-static movements, but can be implemented more reliably and on more systems without the tuned smoothing terms.

Both approaches demonstrated noise insensitivity as well as some robustness to model mismatch (since both controllers utilize a time-varying linearization.)
However, these controllers have yet to be tested with unknown external loads or disturbances.
It is anticipated that such settings may have more impact on the controller with MPC plus inverse statics, since it relies more heavily on open-loop behavior.
In this case, approaches may exist for tuning disturbance rejection, such as increasing pretension in the reference input trajectory via eqn. (\ref{eq:forcedensityrigid_opt_mintension}).






\subsection{Limitations Of Comparison}\label{sec:limitations}

The results provided here compare the top vertebra of the 3D model to the single vertebra in the 2D model.
This comparison is chosen to demonstrate the largest errors of each simulation.
Thus, Fig. \ref{fig:topvert} and \ref{fig:2d_vert_tracking} represent the same geometry of state trajectory, but do not represent the exact same system model.

Though the controller with inverse statics optimization is prototyped in a reduced-order version of the spine, the formulation is general enough to be applied to a multiple-vertebra, 3D spine.
However, such simulations have not been implemented, and as such, it is unknown if some combination of both optimization problems in Sec. \ref{sec:ctler_smoothing} and \ref{sec:ctlr_invstat} may still be required for the higher-dimensional system.

\section{Conclusion}\label{sec:conclusion}


This work contributes two controller formulations and one inverse statics re-formulation for tensegrity robots, as well as simulations showing their efficacy on two models of tensegrity spines.
The second controller, which combines model-predictive control and an inverse statics optimization problem, proposes a new architecture for addressing computational tractability and tuning requirements in tensegrity robots such as these.
The two controllers have different benefits, with higher performance of the MPC plus smoothing formulation compared against the the lower tuning complexity of the MPC plus inverse statics formulation.
The MPC plus inverse statics controller shows tracking performance and noise insensitivity appropriate for use in quasi-static motions of these robots, and is shown to be sufficiently general to apply to different tensegrity spines with no tuning required.

Future work will focus in two areas.
First, performance improvements are needed.
In addition to the computational aspects mentioned above, better tracking may be achieved using inverse dynamics instead of inverse statics solutions.
Using higher-fidelity models, or more sophisticated numerical techniques, may allow for a lower-frequency controller to show good tracking performance.



In addition, hardware experiments using such a lower-frequency controller will be conducted in future work.
Significant mechanical design challenges remain before an appropriate physical prototype can be constructed, particularly with actuation (the dimension of $\bu$) and sensing (since this work uses state feedback.)
Work is ongoing in each area.

\appendix


\subsection{Spine Geometry}\label{appendix:eqns_of_motion}

For the vertebra models in Sec. \ref{sec:dynamics_section}, the local frames of each node from Fig. \ref{fig:topology_3D} and \ref{fig:topology_2D} are, in centimeters,


\begin{align}
\begin{bmatrix}
   \ba_1 \; \;  \ba_2 \; \; \ba_3 \; \; \ba_4 \; \; \ba_5 
\end{bmatrix}
& =
\begin{bmatrix}
    0 & 13 & -13 & 0  & 0 \\
    0 & 0  & 0   & 13 & -13 \\
    0 & -7.5 & -7.5 & 7.5 & 7.5
\end{bmatrix} \\
\begin{bmatrix}
    \ba_1 \quad \ba_2 \quad \ba_3 \quad \ba_4
\end{bmatrix}
& =
\begin{bmatrix}
    0 & 13   & -13   & 0 \\
    0 & -7.5  & -7.5 & 7.5 
\end{bmatrix}.
\end{align}
\vspace{0.1cm}

\noindent The mass at each node is assigned to evenly distribute the $m=0.13$ kg mass of each vertebra.

The larger, differently-shaped 2D vertebra considered in Sec. \ref{sec:different_spines} had $m=0.2$ kg and nodes at
\begin{equation}\label{eq:2D_topology_larger}
\begin{bmatrix}
    \ba_1 \quad \ba_2 \quad \ba_3 \quad \ba_4 
    \end{bmatrix}
=
\begin{bmatrix}
    0 & 20   & -20   & 0 \\
    0 & -20  & -20 & 20 
\end{bmatrix}.
\end{equation}






\subsection{Spine Kinematics and Dynamics}

The system state $\bxi$ parameterizes the position of each point mass within a vertebra.
For a local frame of particle positions $\ba_k$ in vertebra $j$, the particle's position in the global frame is $\bb_{kj}$ as in

\vspace{-0.2cm}
\begin{equation*}
    \bb_{kj}(\bxi) = \bR_j^{\phi}(\bxi) \bR_j^{\gamma}(\bxi) \bR_j^{\theta}(\bxi) \ba_k + \br_j(\bxi),
\end{equation*}

\noindent with the vertebra's center of mass $\br_j$ and rotation matrices $\bR_j$ a function of the generalized coordinates.
The 2D model removes the $y, \theta, \phi$ coordinates, but is otherwise expressed in the same manner.




The continuous-time function $\bg(\bxi, \bu)$ can be symbolically solved by considering $\bb_{kj}$ as a system of particles. 
These models have $J$ vertebrae and $K$ point masses per vertebra.
Lagrange's equations were used to express the dynamics of the system.
With the particles' total kinetic energy $T$, gravitational potential energy $U$, and Lagrangian $L=T-U$, 




\vspace{-0.2cm}
\begin{equation}
    \frac{d}{dt}\frac{\partial L}{\partial \dot \xi_i} - \frac{\partial L}{\partial \xi_i}  = \sum_{j=1}^{J} \sum_{k=1}^K \bF_{kj} \cdot \frac{\partial \bb_{kj}}{\partial \xi_i}, \quad i=1...6J \label{eq:LagrangeMain},
\end{equation}

\noindent The cable force $\bF_{kj}$ uses the spring constant $k=2000 \frac{N}{m}$ and damping constant $c=100 \frac{Ns}{m}$ for eqn. (\ref{eq:springcable}).
The right-hand side and left-hand side of (\ref{eq:LagrangeMain}) are solved symbolically, then equated to solve for $\bg( \bxi, \bu )$, noting that $\dot \xi_i = \xi_{i+6J}$.
Results provided in the accompanying software$^{3}$.

\subsection{Spine State Trajectory}\label{appendix:state_traj}

The state trajectory $\bar \bxi$ for the spine models discussed in Sec. \ref{sec:dynamics_section} separates the vertebrae by 10cm vertically in their starting positions, as per \cite{Sabelhaus2015,Sabelhaus2017}. 
For vertebra $j=1 \hdots J$, 

\vspace{-0.2cm}
\begin{equation}
\bar{z}_{j}(0) = 0.1j \; .
\end{equation}

\noindent These initial heights also define the radius of the rotation: $r_j = \bar{z}_j(0)$.
Consequently, the reference positions of each vertebra over time, $\bar{x}_{j}(t)$ and $\bar{z}_{j}(t)$, are:

\vspace{-0.3cm}
\begin{equation}
\bar{x}_{j}(t) = r_{j} \sin( \beta_j(t)), \quad 
\bar{z}_{j}(t) = r_{j} \cos( \beta_j(t)).
\end{equation}

\noindent In addition, the desired rotation $\bar{\gamma}_j(t)$ of each vertebra about its inertial $Y$-axis is defined to be the same as the sweep angle $\beta_j(t)$ for that vertebra,

\vspace{-0.3cm}
\begin{equation}
    \bar{\gamma}_{j}(t) = \beta_j(t).
\end{equation}

\noindent The maximum sweep angles for each vertebra are the following.
The 2D model with one vertebra only uses $\beta^{max}_1$.


\vspace{-0.4cm}
\begin{equation}
[\beta^{max}_1, \quad \beta^{max}_2, \quad \beta^{max}_3 ] = 
\Bigg[ \frac{\pi}{16}, \quad \frac{\pi}{12}, \quad \frac{\pi}{8} \Bigg].
\end{equation}



For the larger, different spine in Sec. \ref{sec:different_spines}, the adjusted $\bar \bxi$ uses the same $\beta_1^{max}$ but a height of $\bar z_1(0) = 0.3$ m.

\subsection{Simulation Noise Model}\label{appendix:noise_model}

Process noise is implemented by adding a sample from a normally-distributed random variable to the system dynamics during the simulation.
For example, 


\vspace{0.05cm}
\begin{equation}
    \bxi_{t+1} = \bxi_{t|t} +   \bg(\bxi_{t|t}, \bu^*_{t|t}) (\Delta t) + \bE \bm{\epsilon}_t,
\end{equation}

\noindent where $\bm{\epsilon}_t$ is a sample drawn from $\bm{\epsilon} \sim \mathcal{N}(\bzero, \bI)$ at time $t$.
The weighting matrix $\bE$ scales the variance of the random variable, and is given in the accompanying software\footnote{https://github.com/BerkeleyExpertSystemTechnologiesLab/ultra-spine-simulations}.

\section*{Acknowledgment}

Many thanks to those who contributed to earlier versions of this work.
In particular, thanks to Mallory Nation and Ellande Tang for early work on the inverse statics formulation, as well as Abishek K. Akella, Zeerek A. Ahmad, and Vytas SunSpiral for their contributions to the earlier conference version of this paper.
Many thanks to Jeffrey Friesen, Kyunam Kim, Francesco Borrelli, and Andy Packard for feedback on this research.




\bibliographystyle{IEEEtran_nourl}
\bibliography{IEEEabrv,bibliographies/library.bib,bibliographies/IEEEbstctl_nourl.bib}

%






\vspace{-0.3cm}
\begin{IEEEbiography}
[{\includegraphics[width=1in,height=1.25in,clip,keepaspectratio]{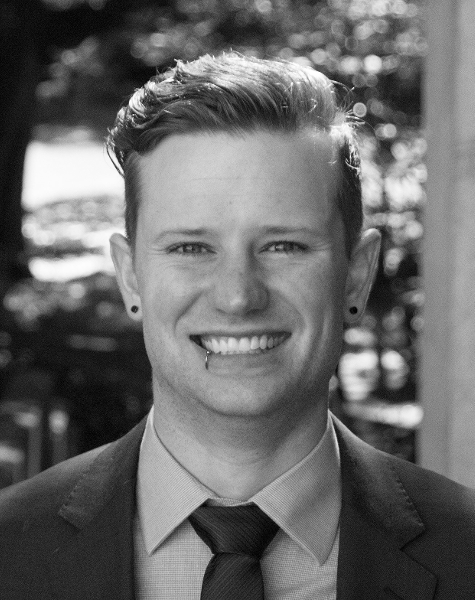}}]{Andrew P. Sabelhaus} received the B.S. degree in mechanical engineering from the University of Maryland, College Park, MD, USA in 2012 and the M.S. degree in mechanical engineering from the University of California Berkeley, CA, USA in 2014, where he is currently pursing a Ph.D. degree.

He has been a National Science Foundation Graduate Research Fellow while at UC Berkeley, and is currently a NASA Space Technology Research Fellow with NASA Ames Research Center, Moffett Field, CA, USA.
\end{IEEEbiography}

\vspace{-0.6cm}
\begin{IEEEbiography}
[{\includegraphics[width=1in,height=1.25in,clip,keepaspectratio]{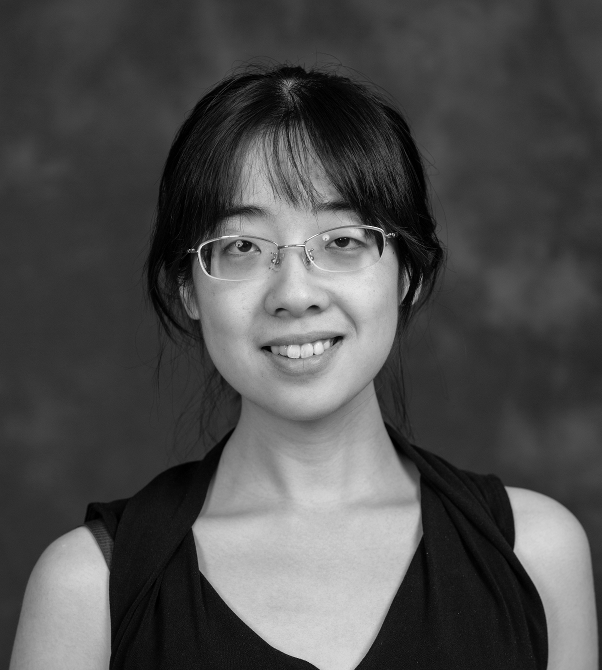}}]{Huajing Zhao} received the B.E. degree in mechanical engineering and automatic manufacturing from Harbin Institute of Technology, China in 2016, and the M.Eng degree in mechanical engineering from the University of California Berkeley, USA in 2017.

She is currently pursuing an M.S. degree in the department of robotics at the University of Michigan, Ann Arbor, MI, USA. She is a member of Automotive Research Center (ARC), and received Best Student Poster Finalist Award at the 2018 ARC Annual Program Review. 
\end{IEEEbiography}

\vspace{-0.6cm}
\begin{IEEEbiography}
[{\includegraphics[width=1in,height=1.25in,clip,keepaspectratio]{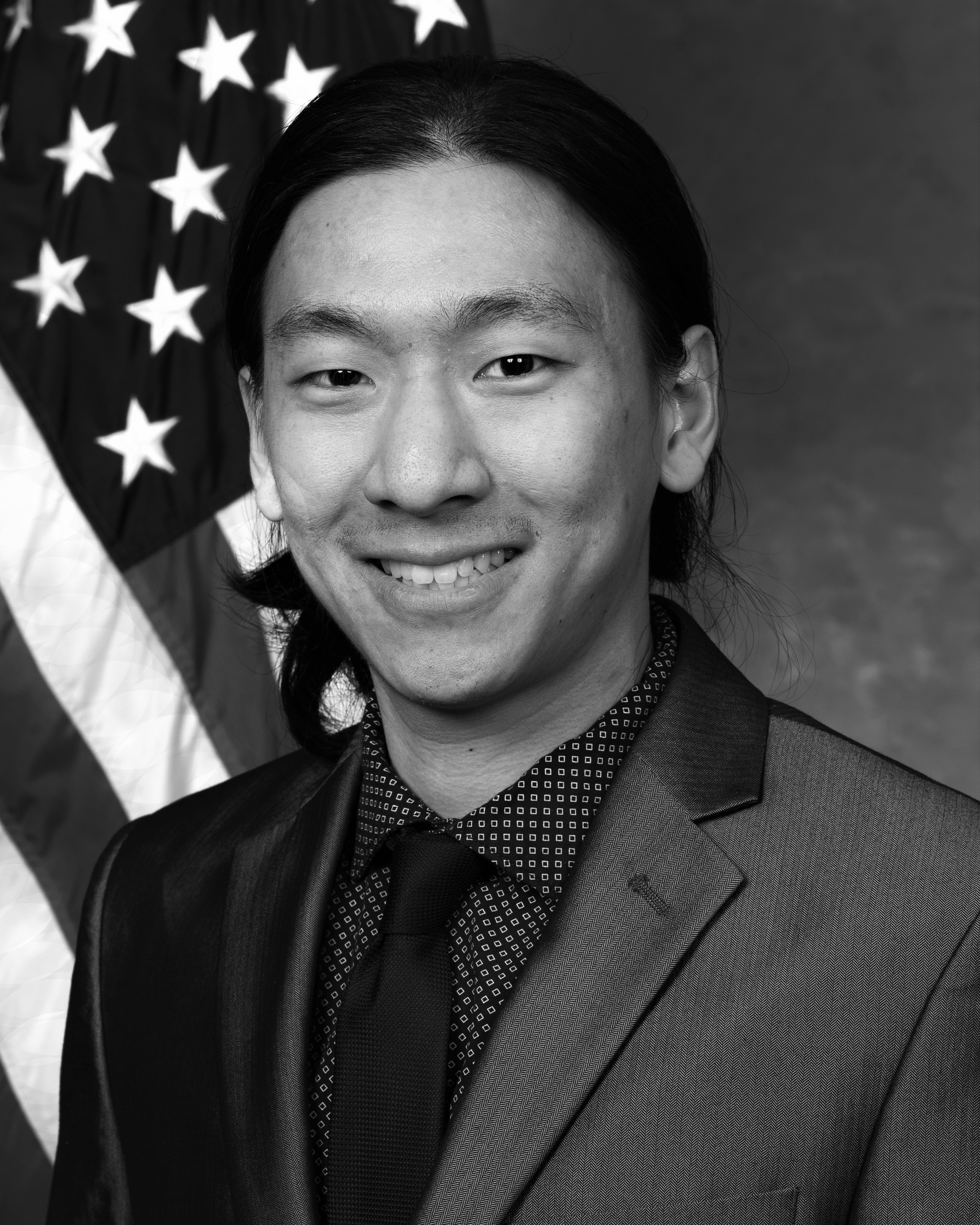}}]{Edward L. Zhu} received his B.S. degree in mechanical engineering from Villanova University, PA, USA in 2015 and his M.S. degree in mechanical engineering from the University of California Berkeley, CA, USA in 2017.

He was a DoD SMART fellow at Berkeley, where his research involved uncertainty estimation and planning for tensegrity robot locomotion. He is currently a Research Scientist at the U.S. Army Research Lab, Vehicle Technology Directorate, Aberdeen, MD, USA. 
\end{IEEEbiography}

\vspace{-0.6cm}
\begin{IEEEbiography}
[{\includegraphics[width=1in,height=1.25in,clip,keepaspectratio]{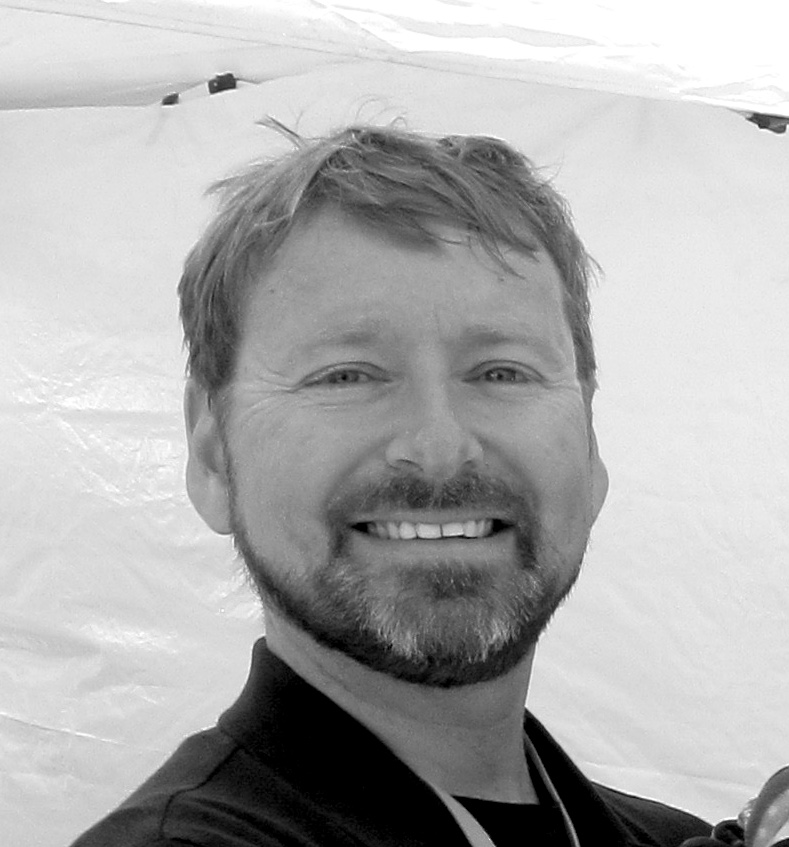}}]{Adrian K. Agogino} received the B.S. degree in computer engineering from the University of California San Diego, La Jolla, CA, USA in 1996, and the M.S. and Ph.D. degrees in electrical and computer engineering from the University of Texas at Austin, TX, USA in 1999 and 2003 respectively.

He is a research scientist at NASA Ames Research Center, Moffett Field, CA, where he has been since 2004. 
He has over 70 publications in the fields of machine learning, soft robotics, rocket analysis, multiagent systems, reinforcement learning, evolutionary systems and visualization of complex systems. He has received awards for publications in both learning and in visualization. 
\end{IEEEbiography}

\vspace{-0.6cm}
\begin{IEEEbiography}
[{\includegraphics[width=1in,height=1.25in,clip,keepaspectratio]{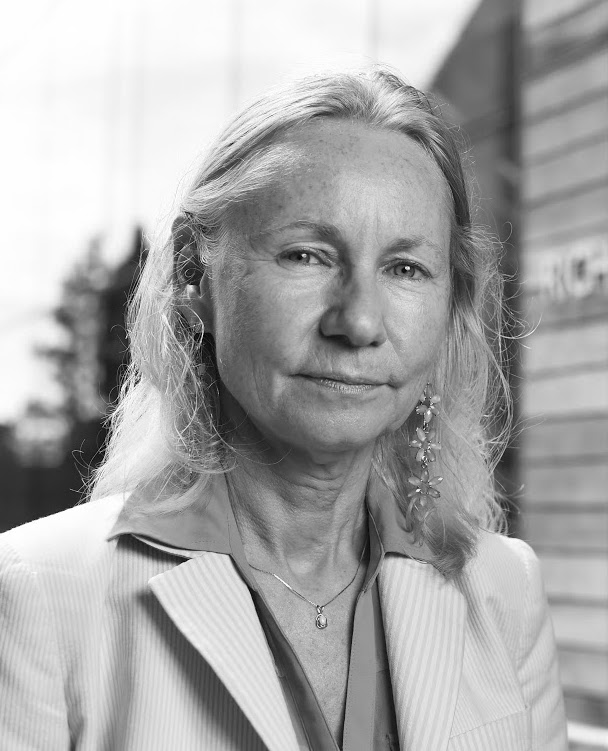}}]{Alice M. Agogino} (SM'14) received her PhD in Management Science \& Engineering from Stanford University, MS in Mechanical Engineering from UC Berkeley and BS in Mechanical Engineering from the University of New Mexico. 

She is currently the Roscoe and Elizabeth Hughes Professor of Mechanical Engineering at UC Berkeley, and serves as Chair of the Graduate Group in Development Engineering as well as Education Director of the Blum Center for Developing Economies. 

Prof. Agogino has authored over 300 peer-reviewed publications, is a member of the National Academy of Engineering (NAE) and has won numerous teaching, mentoring, best paper, and research awards. She has supervised 53 PhD dissertations and 193 MS theses/reports. 
\end{IEEEbiography}

\end{document}